\documentclass[aps,prl,twocolumn,10pt,superscriptaddress]{revtex4-2}

\usepackage{amssymb,amsfonts,amsmath}
\usepackage{times}
\usepackage{graphicx}
\usepackage{color}
\usepackage{float}
\usepackage{cancel}
\usepackage[T1]{fontenc}
\usepackage{lipsum}

\usepackage[latin9]{inputenc}
\usepackage{babel}
\usepackage{bbold}
\usepackage{wasysym}
\usepackage{multirow}
\usepackage{gensymb}
\usepackage[dvipsnames]{xcolor}
\usepackage{pifont}
\usepackage{microtype}
\usepackage{physics}
\usepackage{xcolor}

\usepackage[linktocpage=true,
  colorlinks=true, 
  pdfborder={0 0 0},
  linkcolor=blue,
  citecolor=red,
  filecolor=yellow,
  urlcolor=blue,
  bookmarks,
  pdfauthor={},
]{hyperref}

\newlength{\figurewidth}
\newlength{\pagewidth}
\usepackage{longtable}

\DeclareMathAlphabet\mathbfcal{OMS}{cmsy}{b}{n}
\newcommand{\tc}{$T_\mathrm{c}$}

\begin{document}

\title{Predictions of superconducting properties from chemical bonding interactions}

\author{Francesco Belli}\affiliation{Department of Chemistry, State University of New York at Buffalo, Buffalo, NY 14260-3000, USA}
\author{Trinidad Novoa}\affiliation{Fisika Aplikatua Saila, Gipuzkoako Ingeniaritza Eskola, University of the Basque Country (UPV/EHU), Europa Plaza 1, 20018 Donostia/San Sebasti\'an, Spain}
\affiliation{Centro de F\'isica de Materiales (CFM-MPC), CSIC-UPV/EHU, Manuel de Lardizabal Pasealekua 5, 20018 Donostia/San Sebasti\'an, Spain}
\author{Ion Errea}\affiliation{Fisika Aplikatua Saila, Gipuzkoako Ingeniaritza Eskola, University of the Basque Country (UPV/EHU), Europa Plaza 1, 20018 Donostia/San Sebasti\'an, Spain}
\affiliation{Centro de F\'isica de Materiales (CFM-MPC), CSIC-UPV/EHU, Manuel de Lardizabal Pasealekua 5, 20018 Donostia/San Sebasti\'an, Spain}
\affiliation{Donostia  International Physics Center (DIPC), Manuel de Lardizabal Pasealekua 4, 20018 Donostia/San Sebasti\'an, Spain}
\author{Eva Zurek}\email{ezurek@buffalo.edu}\affiliation{Department of Chemistry, State University of New York at Buffalo, Buffalo, NY 14260-3000, USA}

\begin{abstract}
The need for rapid estimates of superconducting critical temperatures, \tc s,  is becoming increasingly urgent as first-principles and artificial intelligence based methods enable the exploration of ever-larger and more chemically complex materials spaces. In this context, chemical bonding based quantities have emerged as promising descriptors for predicting superconductivity.  To date, these approaches have mainly been developed for hydride superconductors, notably through the networking value, a descriptor based upon the electron localization function, whose topological properties have been found to correlate with \tc . In this perspective, we review and consolidate the method's theory, and we discuss how the estimate of the \tc\ is achievable within a real-space view. We conclude by proposing a framework that enables \tc\ prediction for the broad family of conventional superconductors from easily computable intermediate physical quantities. 
\end{abstract}

\maketitle


\section{Introduction}
Since its discovery in 1911~\cite{Onnes}, superconductivity has fascinated physicists, chemists, and materials scientists, as evidenced by five  Nobel prizes (the first awarded to Onnes in 1913), tens-of-thousands patents filed, and intense ongoing research~\cite{Livas,Zurek:2021k,highentropySC,yao2021superconducting,nickelate,Zurek:2025q}. The potential rewards for finding ``better'' superconductors are immense due to these materials real-world technological uses: as wires conducting electricity without energy loss, powerful and compact magnets in wind turbines and in medical imaging, high precision magnetic measuring devices, and as electronic circuit  qubits for quantum computing.  While ``better'' is often associated with higher superconducting critical temperatures, \tc s,  other properties, including ductility~\cite{ponce} and malleability, ease of synthesis, and ability to sustain a high superconducting critical current or magnetic field~\cite{Geisler:2026a}, can be key factors driving industrial adoption.  The discovery of an easy-to-process room-temperature superconductor would change our lives in ways that cannot be imagined.

\begin{figure*}
    \centering
    \includegraphics[width=\textwidth]{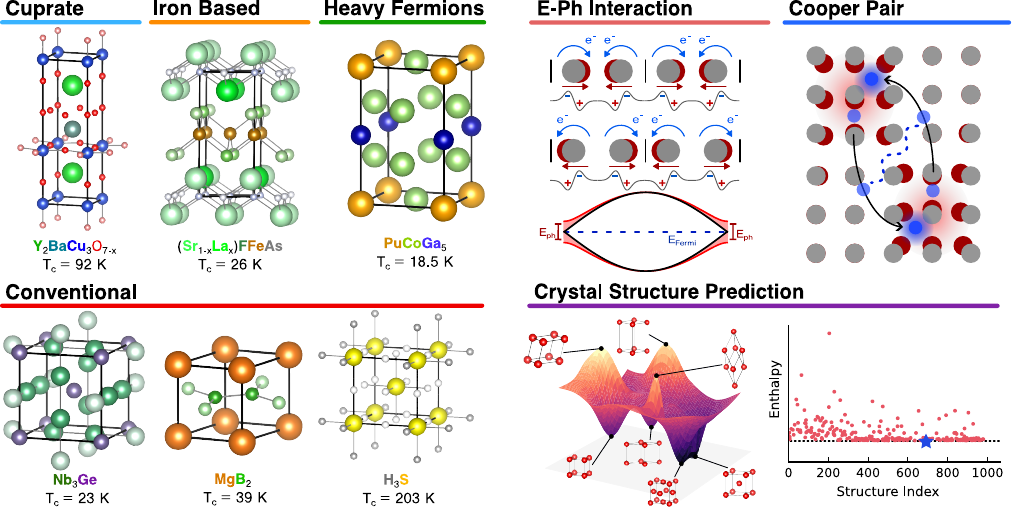}
    \caption{Examples of cuprate, iron based, and heavy fermion (unconventional) superconductors are shown under the cyan, orange and green headers, while examples of conventional electron-phonon (e-ph) mediated systems are shown under the red header. The e-ph mechanism is illustrated under the pink header: atomic vibrations (phonons) generate a displacement-potential that perturbs the electron density, which, in turn, induces a modification of the band structure near the Fermi energy (E$_{\mathrm{Fermi}}$) on the scale of the phonon energy (E$_{\mathrm{ph}}$).  A sketch of the mechanism resulting in the formation of Cooper pairs in conventional superconductors is shown under the blue header; an electron moving through the lattice creates a local distortion, producing a charge imbalance that attracts a second electron. An example of a potential energy surface (PES) is provided under the purple header, highlighting candidates that could be obtained in a crystal structure prediction (CSP) search. In a CSP run, structures are generated and locally optimized until the lowest enthalpy structure (blue star) is found multiple times.}
    \label{fig:overview}
\end{figure*}

Following Onnes' discovery, superconductivity remained an enigma. Early experimental efforts identified only a few simple superconductors, all with relatively low \tc s, such as Pb (7.2~K)~\cite{PhysRev.78.635}, Nb (9.2~K), Nb$_3$Sn (18.3 K)~\cite{PhysRev.95.1435}, and NbTi (10~K)~\cite{scanlan2004superconducting}. A giant leap forward occurred in 1986 with the discovery of cuprate superconductors~\cite{bednorz1986possible}, a group of perovskite-type compounds that contain CuO$_2$ layers. Perhaps the most famous example is YBa$_2$Cu$_3$O$_{7-x}$ (YBCO), the first compound to possess a {\tc} surpassing the boiling point of liquid nitrogen (77~K)~\cite{Chu:1987a}. Prior to the discovery of this class of so-called high-temperature-superconductors (HTSC), expensive cryogens were required to induce the superconducting state. The highest-\tc\ cuprate known is HgBa$_2$Ca$_2$Cu$_3$O$_8$, with \tc= 133~K (164~K under 31~GPa of pressure~\cite{Gao:1994a}, and 151~K after pressure quench to ambient~\cite{Deng:pqp}). As time passed, other classes of superconductors with interesting physics were discovered, but it took $\sim$ 30 years until the \tc\ record was broken again, first by  H$_3$S~\cite{Drozdov:2015a} and  then by LaH$_{10}$~\cite{LaH10_Somayazulu,drozdov2019superconductivity}, with measured \tc s of 200~K at 150~GPa and 250-260~K near 180~GPa, respectively, \emph{en route} towards room temperature. 
However, the brittleness of the cuprates and the high pressures required to stabilize the hydrides hinder their widescale adoption within applications. Examples of the idealized atomistic structures of some superconducting compounds are shown in the left hand side of Figure~\ref{fig:overview}.

While these experimental advances unfolded, the theoretical side saw the development of Nobel-prize winning Bardeen-Cooper-Schrieffer (BCS) theory~\cite{PhysRev.108.1175}, which provided the first plausible microscopic mechanism for superconductivity.  Within BCS theory the superconducting state is associated with the pairing of electrons (with opposite momentum, and usually opposite spin) via an attractive phonon-mediated interaction capable of overcoming their mutual Coulomb repulsion (Figure~\ref{fig:overview}, top right). The resulting Cooper pairs behave as bosons; they are therefore able to condense into the same ground quantum state. This framework was later formalized in Migdal-Eliashberg theory~\cite{migdal1958interaction,eliashberg1960interactions,eliashberg1961temperature} by incorporating the Green's function approach developed by Migdal for the electron-phonon (e-ph) interaction in the normal metallic state. Theories deriving from BCS, however, are limited to e-ph mediated (or conventional) superconductivity. They are therefore not able to predict the properties of unconventional superconductors (examples of which are illustrated in Figure \ref{fig:overview} top left) including cuprates~\cite{annurev:/content/journals/10.1146/annurev-conmatphys-070909-104117,negi2024cuprate}, nickelates~\cite{kitatani2020nickelate}, as well as iron-based~\cite{RevModPhys.83.1589} and heavy-fermion~\cite{brison2000heavy} systems, whose superconducting mechanisms are still under debate.

%

The development of Migdal-Eliashberg theory represented an important paradigm shift: for the first time, it enabled the prediction of the superconducting critical temperature directly from first-principles calculations. Once it became possible to calculate {\tc}, experimental results could be used as benchmarks to refine the theory and establish a more solid foundation. These advances paved the way for the investigation of subtle effects that influence superconducting properties including spin-orbit coupling and/or spin fluctuations, which are important for superconductivity in some elemental metals ~\cite{PhysRevB.106.L180501,kawamura2020benchmark}, and the influence of configurational entropy on the \tc\ of disordered systems and alloys~\cite{ferreira2024ab}. Theoretical studies have yielded a deeper understanding of the  dynamic stabilization, structural renormalization, and anomalous trends in \tc\ (including the inverse isotope effect in PdH~\cite{PhysRevLett.111.177002}), for various ambient and high-pressure hydrides, induced by quantum-nuclear-driven anharmonicity~\cite{belli2025efficient,belli2025chemical}. Furthermore, advances in computational methods have enabled investigations of subtle phenomena such as:  anisotropic contributions to the superconducting gap (crucial in layered materials, such as MgB$_2$~\cite{PhysRevB.87.024505}), full-bandwidth description (important for systems with van Hove singularities near the Fermi energy, such as H$_3$S~\cite{lucrezi2024full}), and the addition of higher-order terms to the e-ph vertex expansion (important for hydrides~\cite{mishra2025electron,bianco2023non,Zurek:2026o}). All these developments have resulted in an incredibly precise computational theory able to estimate {\tc} within a few Kelvin of experiments~\cite{mishra2025electron}.

A second advance, important for materials prediction, has been the development of crystal structure prediction (CSP) methods~\cite{Zurek:2014d,Zurek:2020i}. In the last two decades algorithms that can be used to computationally find the (putative) thermodynamically stable, and important metastable structures for a given chemical composition have become widely adopted. Focussing on inorganic systems, and in particular those under high pressure, some of the most well-known CSP methods include random structure searches (e.g.\ the AIRSS code~\cite{pickard2011ab}), particle swarm methods (e.g.\ CALYPSO~\cite{wang2012calypso}), evolutionary algorithms or EAs (e.g.\ USPEX~\cite{oganov2006crystal} or \textsc{XtalOpt}~\cite{lonie2011xtalopt}), along with metadynamics~\cite{Laio:2002a} and minima hopping~\cite{Godecker:2004} algorithms. Though these methods vary in how structures are created or evolved, all aim to explore the low-lying, and therefore chemically sensible, regions of the potential energy surface (PES) of a user defined chemical combination (as illustrated schematically in Figure \ref{fig:overview}, bottom right).

A number of the aforementioned CSP methods have been further refined towards the discovery of materials with specific functionalities~\cite{hajinazar2025xtalopt}. In standard EAs, which aim to find the global minimum in the PES, thermodynamic quantities are used to determine a structure's fitness, which is related to its likelihood to be chosen for procreation. Multi-objective EA schemes, on the other hand, allow for the incorporation of additional criteria including physical properties (e.g.\ hardness~\cite{Zurek:2019b}), electronic structure (e.g.\ degree of electride character~\cite{Zurek:2025p}) or structural information (e.g.\ similarity with an experimental X-ray diffraction pattern~\cite{Zurek:2024j} or desired crystalline features~\cite{Zurek:2023d}) within the search. These properties can be employed to remove structures from the gene pool (filtration~\cite{Zurek:2023d}), be used within a generalized scalar function~\cite{Zurek:2024e} that merges all objectives to obtain a single measure of fitness for each structure or within the well-known Pareto optimization scheme~\cite{Zurek:2025n}, where optimization proceeds along a multidimensional front of optimal trade-offs.  Instead of minimizing a single variable (e.g., energy), a Pareto search explores the best compromise among multiple relevant properties with the Pareto front representing the optimal hyper-surface balancing all desired quantities.

CSP methods have been widely used in the search for novel inorganic superconducting materials, both to identify the lowest enthalpy structures for a given composition, and to locate all those that comprise the convex hull of thermodynamic stability (focusing on hydrides examples include Refs.~\cite{sanna2024prediction,Wang:2012a,shipley2021,Liu:2017a,Gao:2008a,Zurek:2017c,geng2022structural,yan2020metastable,redington2024predicted,wang2024superconductivity,cataldo2022}). To mitigate the large cost associated with \emph{ab initio} CSP searches, approaches employing machine learning (ML) interatomic potentials for filtering of the initial identified set of structures, followed by DFT relaxation of the most promising species, have been put forward~\cite{salzbrenner2023developments,pickard2022ephemeral}. ML generative models trained on high pressure phases~\cite{luo2024deep},  data-driven techniques~\cite{jiang2024data}, as well as workflows for high-throughput screenings that focus on structural templates~\cite{lucrezi2022silico,GENG2024101443},  have recently been developed. 

The search for superconducting materials via CSP-based multi-objective optimization schemes, which not only optimize thermodynamic stability, but also maximize superconducting properties either by considering features enhancing {\tc}, such as specific geometric motifs or increased density of states (DOS) at the Fermi level, or by embedding the calculation of {\tc} into the multi-objective scheme itself, are the emerging frontier. A quick and reliable way to estimate {\tc} remains the major computational bottleneck towards achieving this goal. While the cost of CSP has been drastically reduced through the aforementioned techniques, calculating the \tc\  across a large number of candidate structures is still prohibitively demanding. 

As a result, significant effort is being devoted to find ways to accelerate \tc\ calculations, for example through simplified Muffin-tin approximations for the potential~\cite{radevych2025rigid}, or via ML models. With regards to ML, deep neural-network models have been developed and employed to screen structures found in public databases for superconductivity, and to predict the magnitude of their \tc ~\cite{wines2024data,wines2023inverse,wines2023high,choudhary2022designing,PhysRevB.101.144505,quinn2022identifying,gibson2503developing,prakash2025guided,cerqueira2024sampling,da2025machine}. Alternative screening approaches have employed ML models trained on the Eliashberg spectral functions of low temperature superconductors~\cite{gibson2025accelerating}, while other studies have derived trends or descriptors based on the analysis of properties that are easy to compute or obtain, such as stoichiometry, electronegativity, atomic mass, composition, space group, volume, DOS, and interatomic distances~\cite{jiang2024data,MA2023101233,PhysRevB.101.144505}.

An alternative path towards estimating {\tc} -- and the one we are concerned with here -- makes use of methods that can quantify chemical bonding in the solid state. In conventional superconductors the onset of superconductivity is governed by the e-ph interaction, which is related to the vibration-induced change in the local overlap of electronic orbitals comprising the occupied conduction bands. For systems with appreciable {\tc}s, this overlap is often highly localized, suggesting that real-space bonding descriptors encompassing the first few neighboring atoms could capture the essential ingredients of the (Cooper) pairing interaction.

Tools for the analysis of chemical bonding and electronic structure such as the Electron Localization Function (ELF)~\cite{savin1997elf}, bond valence functions~\cite{OKeeffe1992}, and bonding analysis tools such as Crystal Orbital Overlap or Hamilton Populations (COOP~\cite{hughbanks1983chains} or COHP~\cite{dronskowski1993crystal}), or the Crystal Orbital Bond Index (COBI~\cite{muler2021crystal}) can be employed to quantify the local orbital overlap. The values associated with these descriptors represent intermediate physical quantities that could be used for \tc\ prediction, and they offer direct information about the bond strengths and related force constants -- information that can help determine the characteristic vibrational energy scale~\cite{naik2023quantum} entering the e-ph coupling. Together these quantities estimate both the electronic and vibrational contributions controlling the magnitude of {\tc}.

In what follows we describe how {\tc} can be estimated in certain classes of superconductors starting from a real-space perspective, thereby providing a theoretical justification for the recently developed models that estimate \tc\ using chemical bonding concepts~\cite{belli2021strong,belli2025refining,PhysRevB.110.174515}. Furthermore, we illustrate how bonding analysis descriptors such as the ELF and COHP can be connected to e-ph theory for conventional superconductivity, and propose ways to employ this approach systematically to accelerate the discovery of new superconductors in conjunction with multi-objective CSP methodologies.

\section{Mathematical Formalism}
Superconductivity arises from the formation of correlated electron pairs extending over length scales ranging from a few, to thousands of nanometers. At first glance, such nonlocal interactions may appear challenging to characterize. Remarkably, \tc\ can be expressed in terms of effective, Fermi-surface-averaged quantities calculated for the primitive unit cell. Advanced theoretical approaches have been developed for estimating {\tc}~\cite{PhysRevB.87.024505}. However, the most widely used expression is the semi-empirical Allen-Dynes modified McMillan (ADM) formula~\cite{PhysRevB.12.905} :
\begin{equation}
    k_\text{B} T_{\mathrm{c}} = \frac{\hbar \omega_{\mathrm{ln}}}{1.2} \mathrm{exp} \left[ \frac{-1.04(1+\lambda)}{\lambda-\mu^*(1+0.62\lambda) } \right],
    \label{eq:McMillan}
\end{equation}
where $k_\text{B}$ is the Boltzmann constant, and $\mu^*$, $\lambda$ and $\omega_{\mathrm{ln}}$ represent the screened single electron repulsive Coulomb interaction, the attractive e-ph coupling constant, and the logarithmic average phonon frequency, respectively, given as
\begin{equation}
    \mu^* = \frac{N_\mathrm{F}\langle V\rangle}{1+N_\mathrm{F}\langle V\rangle\ln\left( \frac{ \varepsilon_{\text{F}} }{\omega_\text{ph}}\right)},
    \label{eq:mustar}
\end{equation}
\begin{equation}
    \lambda = \int_0^{\infty} \frac{\alpha^2\mathrm{F}(\omega)}{\omega} d\omega,
    \label{eq:lambda}
\end{equation}
and,
\begin{equation}
    \omega_{\mathrm{ln}} = \mathrm{exp} \left[ \frac{2}{\lambda} \int_0^{\infty} \frac{ \alpha^2\mathrm{F}(\omega)}{\omega} \mathrm{ln}(\omega) d\omega\right].
    \label{eq:wln}
\end{equation}
In these equations, $N_\mathrm{F}$ represents the electronic DOS at the Fermi level, $\varepsilon_\text{F}$, $\langle V\rangle$ is the average of the Coulomb matrix elements on the Fermi surface, and $\omega_\text{ph}$ is the characteristic phonon frequency of the system. Equation \ref{eq:McMillan} provides a good estimate of \tc\ in the limit of $\lambda \le 1$. For larger values of $\lambda$ this equation underestimates \tc\ as it lacks the strong-coupling and shape corrections included in the less intuitive but more advanced Allen-Dynes approach~\cite{PhysRevB.12.905}. As with many semi-empirical schemes, the strength of the ADM formulation stems from the ability to interpret the system-specific variables, $\lambda$,  $\omega_{\mathrm{ln}}$, and (to a lesser extent) $\mu^*$, in terms of physical properties of the compound.

In Equations \ref{eq:lambda} and \ref{eq:wln} the quantity $\alpha^2\mathrm{F}(\omega)$ represents the Eliashberg spectral function, which describes the scattering  of electronic states near the Fermi energy induced by atomic vibrations and takes the form 
\begin{equation} 
\begin{split}
\label{eq:a2F}
    \alpha^2\mathrm{F}(\omega) = & \frac{1}{N_\mathrm{F}N_qN_k}\sum_{\mathbf{k}\mathbf{q},nm\nu} \left|g_{nm\nu}(\mathbf{k},\mathbf{q}) \right|^2 \cross \\ 
    & \delta(\varepsilon_{n\mathbf{k}}-\varepsilon_{\mathbf{F}})\delta(\varepsilon_{m\mathbf{k+q}}-\varepsilon_{F})\delta(\hbar\omega-\hbar\omega_{\mathbf{q}\nu}).
    \end{split}
\end{equation}
Here 
\begin{equation}
g_{nm\nu}(\mathbf{k},\mathbf{q}) = \bra{\psi_{\mathbf{k}n}} \Delta_{\mathbf{q}\nu} V^\text{KS} \ket{ \psi_{\mathbf{k+ q}m}}
\end{equation}
corresponds to the e-ph matrix elements between a state $\bra{\psi_{\mathbf{k}n}}$ with crystalline momentum $\mathbf{k}$, band index $n$ and energy $\varepsilon_{n\mathbf{k}}$, and a state $\mid\psi_{\mathbf{k+ q}m}\rangle$ with crystalline momentum $\mathbf{k+ q}$, band index $m$ and energy $\varepsilon_{n\mathbf{k}}$. $\Delta_{\mathbf{q}\nu}V^\text{KS}$ is the derivative of the Kohn-Sham potential with respect to an atomic displacement induced by the phonon mode $(\mathbf{q,\nu})$ with frequency $\omega_{\mathbf{q}\nu}$  given as
\begin{equation}
\Delta_{\mathbf{q}\nu} V^\text{KS} = \frac{1}{N_p}\sum_{a\gamma,p}\sqrt{\frac{\hbar}{2M_a\omega_{\mathbf{q}\nu}}}\epsilon_{a\gamma,\nu}(\mathbf{q})e^{i\mathbf{q}\cdot \mathbf{R}_p}\Delta V^\text{KS}_{a\gamma,\mathbf{R}_{p}}.
\label{eq:vks}
\end{equation}
In Equation \ref{eq:vks} $M_a$ is the mass of atom $a$,  $\mathbf{R}_p$ is the lattice vector of the $p$-th unit cell, $\Delta V^\text{KS}_{a\gamma,\mathbf{R}_p}$ is the derivative of the Kohn-Sham potential with respect to the displacement of atom $a$ along direction $\gamma$ in cell $\mathbf{R}_p$, and $\epsilon_{a\gamma,\nu}(\mathbf{q})$ is the phonon eigenvector. Furthermore, $N_k$, $N_q$ and $N_p$ represent the number of points in the \textbf{k}-mesh and \textbf{q}-mesh used in the integrations, and the number of unit cells, respectively. 

At first glance, Equation \ref{eq:a2F} may appear complicated, but it can be understood in terms of Fermi's golden rule, which describes the rates of scattering between quantum states induced by a weak perturbation, averaged over the Fermi surface. Specifically, it describes  transitions between electronic states, $\psi_{\mathbf{k}n} \rightarrow \psi_{\mathbf{k+ q}m}$, induced by the phonon-mediated variation of the Kohn-Sham potential, $ \Delta_{\mathbf{q}\nu} V^\text{KS}$. This perturbation arises from the atomic displacement associated with phonon mode $({\mathbf{q},\nu})$ given by the normalized eigenvector $\epsilon_{a\gamma,\nu}(\mathbf{q})$ with amplitude  $\sqrt{\frac{\hbar}{2M_a\omega_{\mathbf{q}\nu}}}$.

\section{Real Space Interpretation} 
A commonly used approach to evaluate the e-ph matrix elements entering Equation~\ref{eq:a2F} relies on a real-space formulation based on localized functions, obtained by performing a Wannier transformation~\cite{PhysRev.52.191} of the plane-wave Bloch states as:
\begin{equation}
    \ket{\phi_{n,\mathbf{R}_p}} = \frac{1}{N_k} \sum_{n'\mathbf{k}} e^{-i\mathbf{k}\cdot \mathbf{R}_p} U_{nn',\mathbf{k}} \ket{\psi_{\mathbf{k}n'}}.
\end{equation}
Here, \(U_{nn',\mathbf{k}}\) is a \(\mathbf{k}\)-dependent unitary transformation, and \(\phi_{n,\mathbf{R}_p}\) is a localized function centered at cell \(\mathbf{R}_p\). The index $n$ labels the Wannier function, encoding both its atomic site and orbital character. Using this transformation, along with Equations 6 and 7, the e-ph matrix elements in reciprocal space can be expressed as:
\begin{equation}
\label{eq:a2f_RS}
\begin{split}
    g_{nm\nu}(\mathbf{k},\mathbf{q}) =& \sum_{\substack{\mathbf{R}_p,\mathbf{R}_{p'}\\n'm'a\gamma}} e^{i(\mathbf{k}\cdot\mathbf{R}_p + \mathbf{q}\cdot\mathbf{R}_{p'})} \sqrt{\frac{\hbar}{2 M_a \omega_{\mathbf{q}\nu}}} \epsilon_{a\gamma,\nu}(\mathbf{q}) \\
   & \times U_{n'n,\mathbf{k+q}}\, g_{n'm'a\gamma}(\mathbf{R}_p,\mathbf{R}_{p'}) \, U_{mm',\mathbf{k}}^{\dagger},
\end{split}    
\end{equation} 
where 
\begin{equation}
g_{n'm'a\gamma}(\mathbf{R}_p, \mathbf{R}_{p'}) = \langle \phi_{n',0} \mid \Delta V^{\text{KS}}_{a\gamma,\mathbf{R}_{p'}} \mid \phi_{m',\mathbf{R}_p} \rangle
\label{eq:g_RS}
\end{equation}
are the real-space counterparts of the e-ph matrix elements. 

\begin{figure}[h]
	\centering
	\includegraphics[width=\columnwidth]{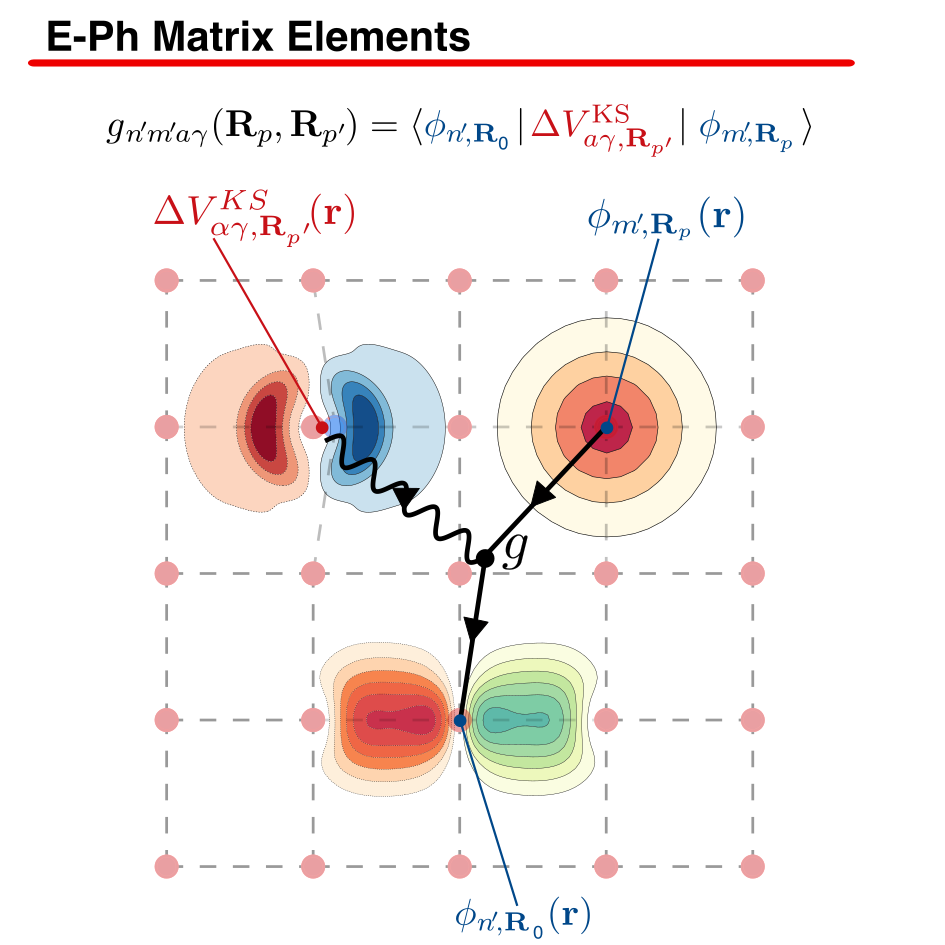}
	\caption{Schematic representation of the quantities entering Equation~\ref{eq:g_RS} for the calculation of the real space electron-phonon (e-ph) matrix elements on a 2-dimensional plane. The black lines describe the e-ph scattering processes from $\phi_{m',\mathbf{R}_p}$ to  $\phi_{n',\mathbf{R}_0}$ induced by the atomic displacement $\Delta V^{\text{KS}}_{a\gamma,\mathbf{R}_{p'}}$. The localized (red-yellow-green-blue) functions describe the electronic states in the proximity of atomic sites (red dots), while the red and blue functions describe the perturbation induced in the potential due to an atomic displacement. }
	\label{fig:A}
\end{figure}

A sketch of the quantities entering Equation~\ref{eq:g_RS} is shown in Figure~\ref{fig:A}, providing an intuitive real-space picture of the e-ph interaction. Here, $g_{n'm'a\gamma}(\mathbf{R}_p, \mathbf{R}_{p'})$ is the matrix element that describes the coupling between two localized electronic states, $\phi_{m',\mathbf{R}_p}$ and $\phi_{n',\mathbf{R}_0}$, induced by a lattice displacement associated with a particular phonon mode. The change in the Kohn-Sham potential, $\Delta V^{\text{KS}}_{a\gamma,\mathbf{R}_{p'}}$, resulting from this vibration mediates the coupling between these two electronic states, and the  strength of this coupling determines the probability of scattering between them via Fermi's golden rule. 

A key advantage of the Wannier formulation lies in the spatial localization of the basis functions. For Equation \ref{eq:g_RS} to be nonzero,  the two Wannier functions $\phi_{n',0}$ and $\phi_{m',\mathbf{R}_p}$ must have a non-negligible spatial overlap.  In practice, the function $\phi$ decays rapidly with distance~\cite{PhysRevB.76.165108}, implying that contributions are significant when the orbitals involved in the scattering and potential variation are located within a few neighboring atomic sites. This locality implies that the e-ph interaction is short-ranged in real space.

%
%

%
\begin{figure*}
	\centering
	\includegraphics[width=\textwidth]{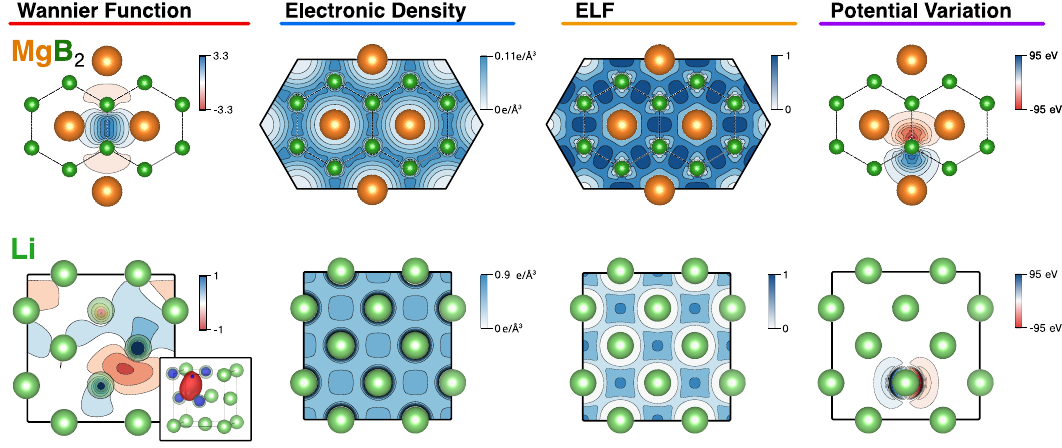}
	\caption{Contour plots and/or isosurfaces for a characteristic maximally localized Wannier function or MLWF (red header), valence electron density (blue header), ELF (yellow header), and potential variation (purple header) for MgB$_2$ (top) and fcc-lithium at 35 GPa (bottom). An isosurface of one of the MLWFs computed for lithium is shown in the inset. Notice the topological resemblance of the illustrated MLWF, electron density and ELF computed for MgB$_2$. In lithium, on the other hand, the region of space where the contour values for the MLWF are highest do not correspond to regions that possess large values of the electron density or ELF.}
	\label{fig:B}
\end{figure*}

Equation~\ref{eq:g_RS} can be interpreted as an overlap between atom or bond-centered localized orbitals, modulated by a potential variation induced by an atomic displacement. However, Wannier functions are not unique and their functional forms are determined by the choice of the unitary transformation, $U_{nn',\mathbf{k}}$. Maximally localized Wannier functions (MLWFs) proposed by Marzari and Vanderbilt~\cite{RevModPhys.84.1419} are among the most commonly used. In MLWFs $U_{nn',\mathbf{k}}$ is chosen to minimize a cost function measuring the orbital spread, equivalent to the Foster-Boys criterion from quantum chemistry \cite{RevModPhys.32.300}. Alternatively, Wannier functions may be generated by minimizing weighted sums of both the spatial and energy variance~\cite{16r3-hjc7}, projecting onto localized atomic orbitals~\cite{PhysRevB.75.045110}, or via symmetrical orthonormalization of order-$N$ Muffin-Tin Orbitals ($N$MTOs), which do not rely on the plane wave orbital projection~\cite{zurek2005muffin}. The resulting shape and extent of localization depends upon the method used; nevertheless an important requirement is that the function retains most of the spatial weight in regions characterized by significant electron density~\cite{RevModPhys.84.1419}. 


The conventional superconductor with the highest \tc\ at 1~atmosphere (39~K) is MgB$_2$ (whose unit cell is shown in Figure \ref{fig:overview}). This compound can be formally described as Mg$^{2+}$(B$^-$)$_2$, adopting a layered structure analogous to it's isoelectronic cousin, graphite. The boron atoms in MgB$_2$ form a honeycomb lattice with magnesium atoms located above and below the center of the hexagons. Analogous to graphite, each boron atom's in-plane $p_x$- and $p_y$-orbitals hybridize with the $s$-orbital forming a set of three $\sigma$-bands, in addition to a $\pi$-band from the out-of-plane $p_z$-orbital~\cite{PhysRevLett.86.4366}. However, unlike graphite, MgB$_2$ has an intricate Fermi surface because the electrostatic potential from Mg$^{2+}$ lowers the energy of the $\pi$-bands relative to the $\sigma$-bands, such that both cross the Fermi level. As a result both bands contribute towards superconductivity, rendering MgB$_2$ a prototypical example of a multiband superconductor, with the stronger e-ph coupling arising from the B-B stretching mode within the honeycomb layers. The MLWFs associated with the $\sigma$-bonds within this layer (one of which is illustrated in Figure \ref{fig:B}) resemble the localized $\sigma$ molecular orbitals computed for benzene~\cite{jochen}. Contour plots of the valence electron density, and ELF of MgB$_2$ (also shown in Figure \ref{fig:B}) clearly possess features connecting the B-B bonds. Thus, plots of the MLWF, electron density and ELF describing the B-B $\sigma$-bonds within MgB$_2$, look qualitatively similar, and one could say that they are topologically consistent. The e-ph interaction associated with the vibrations of the $\sigma$-bonds, $\lambda_\sigma$, is larger than $\lambda_\pi$.

Near 7.5~GPa and at low temperature lithium's body-centered-cubic (bcc) phase transitions to a face-centered-cubic (fcc) structure, whose atom-projected electronic DOS is characterized by $sp$ hybridization. Pressure induces superconductivity within lithium, with the \tc\ of the fcc phase computed to be 13~K at 35~GPa, in good agreement with experiment~\cite{Zurek:2024a}. Despite its nearly-free-electron-like DOS, quantum mechanical calculations reveal that fcc Li possesses non-nuclear maxima (NNM) found within octahedral interstitial sites~\cite{naumov2015chemical,Zurek:2024a}. In fact, at this pressure fcc lithium is an electride -- a compound where the electrons, which are localized in interstitial regions, act as anions~\cite{Zurek:2024p}. Because generating Wannier Functions that span only the occupied bands in a metal can require the construction of large supercells that result in band back-folding~\cite{zurek2005muffin}, we generated MLWFs that spanned the four lowest energy bands, extending up to 6~eV above the Fermi level, by placing $s$-type orbitals on the lithium atoms as initial guesses. Four qualitatively similar MLWFs were obtained; one is illustrated in Figure \ref{fig:B}. 

The generated MLWFs is overall more delocalized but posses strong localization on the lithium atoms with additional character in the interstitial regions where the NNM are found. In fact, they resemble the NNM-centered WF (generated by the $N$MTO method) for another high-pressure-electride, Na-$hP4$, which was interpreted as originating from the overlap of Na-centered $spd$ hybrids resulting in the formation of weak multi-centered bonds~\cite{racioppi2023electride}. A similar interpretation is likely relevant for fcc-Li, where the interstitial electron localization originates from the overlap of Li $sp$ hybrids within multi-centered bonds. Comparison of the MLWFs obtained for Li with plots of the electron density and the ELF all show similarities, though low isosurfaces in the density are required to observe the contributions of the interstitial regions, which are much more pronounced in plots of the ELF that coincide with the location of the NNM. Despite the low electron population in the NNM, the density in these regions is strongly localized, resulting in high ELF values at the NNM sites. 

The different topology of the MLWFs in MgB$_2$ and Li are noteworthy and provide insights on the type of bonding interactions featured in the systems. In MgB$_2$, the Wannier functions are characteristic of localized bonding $\sigma$ orbitals~\cite{jochen}. In contrast, the MLWFs spanning the valence and some unoccupied states in lithium are significantly more delocalized, reflecting the metallic nature of the system.

\section{Site overlap and networking}
Equation~\ref{eq:g_RS} includes the derivative of the Kohn-Sham potential induced by an atomic displacement. The magnitude, shape, and spatial extent of this potential variation  is determined by the underlying electronic structure and bonding topology of the system. The right hand panel of Figure \ref{fig:B} illustrates  $\Delta V^{\text{KS}}_{a\gamma,\mathbf{R}_{p'}}$ calculated for the displacement of a single atom along a Cartesian direction chosen so as to maximize the overlap of the derivative of the potential with regions of the electron density arising from states near the Fermi level. In MgB$_2$ the $\sigma$-bands, which can be described by the bond-centered MLWF in Figure \ref{fig:B}, form hole-pockets resulting in two nearly cylindrical Fermi surfaces extending along the $\Gamma-A$ direction. The potential derivative in this system decays slowly, retaining non-negligible values in the MLWF centers, reflecting the localized and directional nature of the covalent bonding and the large interaction between the potential and the electronic states. 

In the $sp$ metal lithium, on the other hand, the states near the Fermi level correspond to the maxima in the electron density located in the octahedral interstitial regions characterised by high ELF values that are not well described by the generated MLWFs (Figure \ref{fig:B}). In contrast to the results obtained for MgB$_2$, the perturbation decays rapidly within a distance comparable to the core radius, consistent with lithium's nearly free-electron character. This fast decay is evident in Figure~\ref{fig:B}, where the bulk of the potential variation is localized around the lithium atom, exhibiting minimal overlap with the electronic density residing in the octahedral interstitial regions.

The previous discussion illustrates that covalently bonded metals are characterized by factors favoring strong e-ph coupling in conventional superconductors: electronic states near the Fermi energy that are well-described by atom- and/or bond-centered Wannier functions, and that overlap with regions of significant potential variation, thereby maximizing the e-ph coupling matrix elements described in Equation \ref{eq:g_RS}.  These directional bonding/antibonding states are highly sensitive to atomic displacements, so even small vibrations generate strong variations in the electronic structure and, consequently, large e-ph matrix elements. This phenomenon, first discussed by Cohen in Ref.~\cite{cohen1994comments}, is well documented for hydrogen-rich superconductors where covalently bonded networks, strengthened by pressure and electron donation, are responsible for values of {\tc} close to room temperature~\cite{belli2021strong,belli2025refining}, in stark contrast to the {\tc} achievable in simple $sp$-metals and transition metals.

\begin{figure*}
    \centering
    \includegraphics[scale=1]{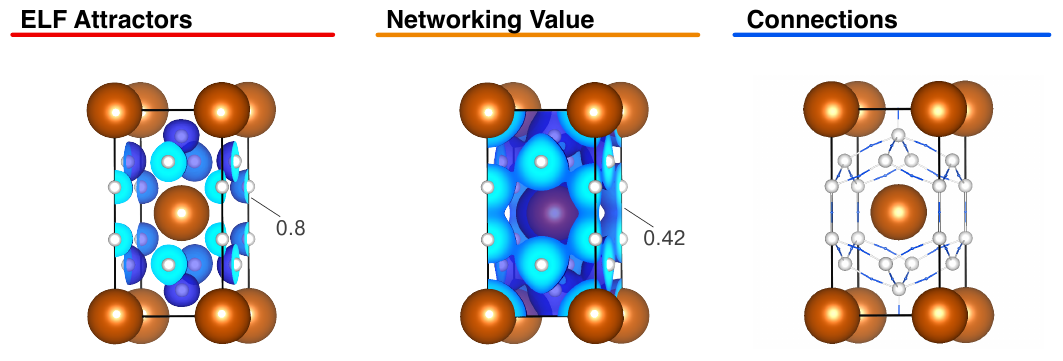}
    \caption{A schematic illustration of the procedure used to obtain the networking value, $\Phi_\text{NV}$, for $I4/mmm$ YH$_4$. First the electron localization function (ELF) attractors are identified (left), next the value of the ELF is lowered until the basins connect and form an extended 3-dimensional network throughout the crystal (middle). The explicit connections between the ELF attractors that form the network are shown in the last panel (right).}
    \label{fig:C}
\end{figure*}

Because the e-ph interaction can be viewed as a local bonding-driven phenomenon, descriptors quantifying the strength and spatial extent of chemical bonding should be able to provide effective proxies for the magnitude of the dominant e--ph matrix elements, without requiring an explicit evaluation of the potential derivative. Potentially suitable descriptors include bond valence functions, the integrated COHP and COBI (ICOHP and ICOBI), and the ELF, which characterize the strength of chemical bonding or the degree of local electron sharing.

In systems with strongly localized covalent bonds, the topology of the ELF attractors identify regions of localized bonding electrons, and the critical ELF value along the pathway connecting neighboring attractors provides a measure of the local orbital overlap between the bonded atoms. The ``networking value'' has been defined as the largest ELF isovalue, $\Phi_\text{NV}$, for which the corresponding isosurface forms a three-dimensional network spanning the crystal (Figure \ref{fig:C}). For hydride superconductors it has been shown that combining $\Phi_\text{NV}$ with  the hydrogen orbital-derived density of states at the Fermi level, $H_{\mathrm{DOS}}$, and the hydrogen atomic fraction, $H_\text{frac}$, can provide an estimate of the superconducting critical temperature via
\begin{equation}
    \label{eq:Tc_estime}
    T_{\mathrm{c}} = 750\Phi_\text{NV} H_\text{frac} H_{\mathrm{DOS}}^{1/3} -85,
\end{equation}
with a root-mean-square error of $\sim$60~K~\cite{belli2021strong}. The predictive power of this method is illustrated for a set of hydride superconductors in Figure \ref{fig:D}(a). Although the quantitative agreement is imperfect, the results demonstrate that chemically motivated descriptors derived from the ELF can capture the dominant chemical factors governing {\tc}. 

Improvements to the original fit~\cite{belli2021strong} have been proposed by refitting using extended datasets~\cite{belli2025refining}, development of automated methods for determining the networking value~\cite{D4SC04465G}, introduction of the molecularity index to account for the presence of H$_2$ units~\cite{PhysRevB.110.174515}, and incorporation of a descriptor that includes the derivatives of the KS potential with respect to the displacement of each atom along the Cartesian directions~\cite{novoa2025real}. However, it has recently been shown that (some of) the revised fits are still unable to resolve large {\tc} differences within a given family of multi-metal hydrides~\cite{Zurek:2026e}. 

Although the predictive accuracy of the aforementioned fits could potentially be improved by introducing further chemically-inspired parameters, 
the intrinsic limitations of ELF-based methods must be noted.
For example, the ELF is constructed from all occupied electronic states, not just those  near the Fermi level. Secondly, the current approaches assume that {\tc} is predominantly driven by hydrogen-derived states; while this assumption is sensible the inability of the RS5~\cite{belli2025refining} fit to predict the 150~K variations in {\tc} for different YCaH$_{12}$ structures suggests that it is insufficient~\cite{Zurek:2026e}. Thirdly, as in the case of electride fcc-lithium, the magnitude of the ELF attractors does not reflect the amount of electron density found within the NNM basins. 
Furthermore, the planewave based codes typically employed for extended systems calculate the atom projected DOS by associating a spherical volume to each atom, and the results of the projection are sensitive to the size of the sphere(s) employed, especially for systems with interstitial voids~\cite{Zurek:2023e} and molecular motifs (such as in Mg$_2$IrH$_6$~\cite{wang2024mg2irh6,sanna2024prediction,PhysRevLett.132.166001}). In addition, the ELF-based approach remains largely phenomenological and lacks a rigorous theoretical foundation. 

In what follows, we derive a real-space framework that provides the missing microscopic justification for ELF-based descriptors, explains both their successes and failures, and enables the construction of more accurate chemically motivated descriptors based on quantities such as the ICOHP and ICOBI.

\section{Theoretical framework.}
\begin{figure}
    \centering
    \includegraphics[width=0.9\columnwidth]{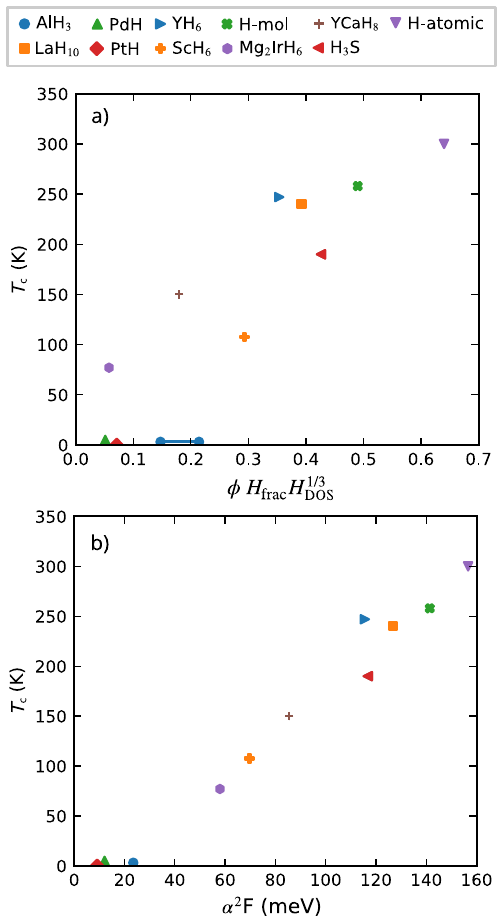}
    \caption{(a) The prefactor of the slope in Equation \ref{eq:Tc_estime} obtained by calculating the networking value ($\Phi_\text{NV}$), the projection of the DOS onto the hydrogen orbitals at the Fermi energy ($H_{\mathrm{DOS}}$) and the fraction of hydrogen atoms ($H_\mathrm{frac}$) comprising the compound. (b) The correlation of the {\tc} with the integral of the Eliashberg spectral function, illustrating the relationship suggested by Equation \ref{eq:LC}. Both panels include data for AlH$_3$ (150 GPa)~\cite{hou2021strong}, PdH (ambient)~\cite{PhysRevLett.111.177002}, YH$_6$ (150 GPa)~\cite{Troyan}, YCaH$_8$(150 GPa)~\cite{zhao2025prediction}, H-$I4_1/amd$ (atomic, 500 GPa)~\cite{atomicH}, H-$Cmca$-4 (molecular, 450 GPa)~\cite{H_mol}, LaH$_{10}$ (250 GPa)~\cite{errea2020quantum}, PtH (100 GPa)~\cite{ErreaSSCHA}, ScH$_6$-$P3_6/mmc$ (140 GPa)~\cite{Hou3}, Mg$_2$IrH$_6$ (ambient)~\cite{PhysRevLett.132.166001}, and H$_3$S-$Im\bar{3}m$ (150 GPa)~\cite{errea2015high,mishra2025electron}. The horizontal line in panel (a) between the blue AlH$_3$ points illustrates variations in the prefactor arising from the choice of density functional.}
    \label{fig:D}
\end{figure}

A natural starting point for developing a formula that can estimate {\tc} from chemical-bonding descriptors is the Leavens and Carbotte expression
\begin{equation}
    T_\mathrm{c} = 0.148 \int_0^\infty \alpha^2\mathrm{F}(\omega)\, d\omega = 0.148 \hspace{1mm} \alpha^2\mathrm{F} .
     \label{eq:LC}
\end{equation}
This expression is based on the assumption that, in the strong-coupling regime ($\lambda > 0.6$), the e--ph interaction is larger than the opposing Coulomb repulsion. Under these conditions, {\tc} is determined primarily by the e--ph interaction strength.

Equation \ref{eq:LC} was originally fit using the properties of low temperature superconductors, but the proposed parametrization still holds for known high temperature conventional superconductors ~\cite{pickett2025compressed}. Furthermore, recent studies have shown that this relationship is more general, holding across a broad range of critical temperatures~\cite{belli2025chemical,pickett2025compressed}. From calculations on a set of hydride superconductors a near-linear relationship between the integral of the Eliashberg spectral function and {\tc} is evident (Figure \ref{fig:D}b).  Although the correlation between the two quantities is not perfect, it is sufficiently strong to allow the Eliashberg-spectral function to serve as a proxy for {\tc}.

With this framework in mind, the explicit form of the spectral function of Equation \ref{eq:a2F} can be factored by separating the contributions associated with each displaced atom, corresponding to the sum over $a$ in Equation \ref{eq:a2f_RS}, and subsequently integrating over the Brillouin zone, 
\begin{equation}
    \label{eq:a2F_core}
    \alpha^2\mathrm{F} \propto N_\mathrm{F}  \sum_{ab} \frac{\langle g_a g_b \rangle}{\sqrt{M_a M_b} \langle \omega \rangle}.
\end{equation}
Here $g_a$ and $g_b$ denote the e--ph matrix elements associated with displacements of atoms $a$ and $b$, averaged over the Fermi surface, while $M_a$ and $M_b$ are the corresponding atomic masses. In deriving Equation~\ref{eq:a2F_core}, it is assumed that the phonon frequencies associated with the different atomic displacements can be represented by a single average phonon frequency, $\langle\omega\rangle$. 
This expression, formally derived for the strong-coupling regime, makes explicit the quantities governing conventional superconductivity: the electronic density of states at the Fermi level, $N_\mathrm{F}$, the atomic masses, the characteristic phonon frequency, and, most importantly, the averaged e--ph matrix elements $\langle g_a g_b\rangle$.

Equation~\ref{eq:a2F_core} also provides a theoretical justification for the networking value approach previously proposed for hydrogen-rich superconductors. As discussed above, it is reasonable to expect that, in these materials, a substantial fraction of the Eliashberg spectral function arises from phonon modes dominated by hydrogen motion. Equation~\ref{eq:a2F_core} can therefore be approximated by retaining the terms due to hydrogen,
\begin{equation}
\alpha^2\mathrm{F} \propto H_\text{DOS}\frac{\langle g_H^2 \rangle H_\mathrm{frac}^{2}}{M_H \langle \omega \rangle}.
\label{eq:sim}
\end{equation}
Here, the factor $H_\mathrm{frac}^2$ accounts for the fraction of atoms contributing to the hydrogen-dominated e-ph coupling, ensuring the correct scaling of the Eliashberg spectral function with system size and composition. Since the hydrogen mass is constant, $M_H$ does not influence trends across different hydrides, while the average phonon frequency $\langle\omega\rangle$ is expected to vary only moderately among chemically and structurally similar compounds. Consequently, the dominant quantities controlling the variation of $\alpha^2\mathrm{F}$ are the hydrogen-projected electronic density of states at the Fermi level, the hydrogen e--ph matrix elements, and the hydrogen fraction.

In this model, the networking value serves as a proxy for $\langle g_H^2 \rangle$, as it reflects the degree of local orbital overlap involving hydrogen atoms. Moreover, because stronger covalent bonding generally leads to both larger orbital overlap and stiffer bonds, the networking value is also expected to correlate with the characteristic phonon frequency. Equation~\ref{eq:sim} captures the essential ingredients already present in Equation~\ref{eq:Tc_estime}, with the latter differing primarily in that the exponents have been empirically optimized to achieve the best quantitative agreement with the {\tc}s computed using DFT.

\section{Generalization of the Method} 
\begin{figure}
    \centering
    \includegraphics[width=\columnwidth]{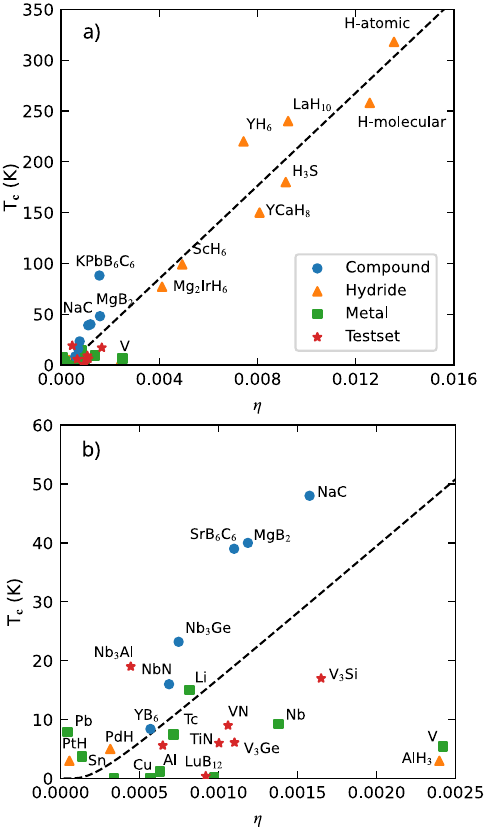}
    \caption{Values of {\tc} (a) with respect to the parameter ($\eta$) proposed in Equation \ref{eq:a2F_estimate}, and (b) a magnified view of this plot in the range from 0-60~K. The correlation is reported for hydrides (yellow triangles), general multi-element compounds (blue circles), atomic metals (green squares), and a test set not used to develop the fit (red stars). The dotted black line represents the best fit obtained through the equation {\tc} $= a/(e^{b/\eta}-1)$.}
    \label{fig:E}
\end{figure}

Equation~\ref{eq:a2F_core} suggests that this approach can be generalized to all conventional superconductors, provided the atom-resolved e--ph matrix elements and the characteristic phonon frequency can be estimated.  A particularly useful quantity for these estimates is the (projected) crystal orbital Hamilton population (pCOHP)~\cite{dronskowski1993crystal}. To calculate this quantity, the wavefunctions obtained from DFT computations are projected onto a localized atomic basis of Slater-type orbitals ($\phi$), with the projection coefficients defined as
\begin{equation}
	T_{nm}(\mathbf{k}) = \braket{\psi_{\mathbf{k}n}}{\phi_{m}}
\end{equation}
Using these projections, the Hamiltonian in the local orbital basis is constructed as
\begin{equation}
	H_{m m'}(\mathbf{k}) = \sum_n \varepsilon_{n\mathbf{k}} T_{nm}^*(\mathbf{k}) T_{nm'}(\mathbf{k}).
\end{equation} 
The pCOHP is then obtained by projecting the Hamiltonian onto the density matrix and integrating over reciprocal space,
\begin{equation}
	\begin{split}
	\mathrm{pCOHP}_{mm'}(\varepsilon) = & \sum_{\mathbf{k}n}\mathbb{R}[T_{nm}^*(\mathbf{k}) T_{nm'}(\mathbf{k})	H_{m m'}(\mathbf{k})] \\
	                          & \cross \delta(\varepsilon_{n\mathbf{k}} -\varepsilon);
	\end{split}
\end{equation} 
it can further be summed over all orbitals belonging to a specific pair of atoms to derive a bond resolved $\mathrm{pCOHP}$
\begin{equation}
	\mathrm{pCOHP}_{ab}(\varepsilon) = \sum_{m \in a} \sum_{m' \in b} \mathrm{pCOHP}_{mm'}(\varepsilon).
\end{equation}
Since the pCOHP is energy resolved, it can be used to provide a direct measure of the bonding (or antibonding) interactions between specific pairs of atoms in a chosen energy window, including the vicinity of the Fermi level where the e-ph interaction is determined.

A recent study reported a correlation between the largest (most negative) value for the integrated pCOHP (ICOHP) for a pair of atoms, corresponding to the strongest bond in a compound, and the crystal's highest phonon frequency~\cite{naik2023quantum}. This chemically intuitive trend suggests that a COHP-based descriptor can serve not only as a measure of the e-ph matrix elements, but also as a proxy for the characteristic phonon frequencies entering Equation~\ref{eq:a2F_core}. Importantly, the pCOHP can be obtained from a single static DFT calculation, requiring only a small fraction of the computational cost of a full phonon calculation, or an even more expensive e-ph coupling calculation.

To develop a fit for \tc, considering its direct dependence on the Eliashberg spectral function (Equation \ref{eq:LC}), the bond resolved $\mathrm{pCOHP}$ can be used to define a volume-normalized mean absolute interaction index for an atom as:
\begin{equation}
K_a = \frac{a_0^3}{V}\frac{\displaystyle \sum_{b} \int_{\varepsilon_1}^{\varepsilon_\mathrm{F}} \left| \mathrm{pCOHP}_{ab}(\varepsilon) \right| d\varepsilon}{2(\varepsilon_\mathrm{F} - \varepsilon_1)}.
\end{equation}
In this equation, $a_0$ is the Bohr radius (introduced to render $K_a$ dimensionless), $\varepsilon_1$ is the lower bound for the integration window below the Fermi energy $\varepsilon_F$, and $V$ is the volume of the cell (introduced to normalize the sum of $K_a$ over all atoms within the cell). The factor of $\frac{1}{2}$ is introduced to prevent double counting, and the absolute value for the $\mathrm{pCOHP}$ is employed to take into account both the bonding and antibonding states, which prevents them from cancelling each other, since they have different signs. The choice of $\varepsilon_1$ is not arbitrary. If chosen to be too close to $\varepsilon_F$, $K_a$ will become artificially small as $\left| \mathrm{pCOHP}_{ab}(\varepsilon) \right|$ decreases due to the sign inversion of the $\mathrm{pCOHP}$ that typically occurs in the proximity of $\varepsilon_F$. Conversely, choosing a too large value of $\varepsilon_1$ results in the inclusion of contributions from electronic states that may not be relevant for superconductivity. In the tests performed here, the optimal choice for $\varepsilon_1$ was found to be on the order of about 0.5~eV.  

$K_a$ is used to construct two approximants for the quantities entering Equation \ref{eq:a2F_core}, which when combined with Equation \ref{eq:LC},  provide a direct measure of \tc\ in the strong coupling regime. The first is the average phonon frequency weighted by the density of states near the Fermi level
\begin{equation}
\langle \omega \rangle N_\mathrm{F} \sim \sum_a K_a \left( \sum_{s \in \text{species}} \frac{M_0}{M_s} \right)^{1/2},
\label{eq:est1}
\end{equation}
where the sum of $K_a$ is over each atom in the unit cell, $M_0$ is the mass of the hydrogen atom introduced to retain the adimensionality, and where the sum over $s$ runs over each distinct chemical element, or species, present in the cell. 
The second quantity provides an estimate for the e-ph matrix elements, considering their dependence on the atomic orbital overlap
\begin{equation}
N_\mathrm{F}^2 \sum_{ab} \frac{\langle g_a g_b \rangle}{\sqrt{M_a M_b}} \sim \left( \sum_a K_a\sqrt{\frac{M_0}{M_a}} \right)^2.
\label{eq:est2}
\end{equation}
By substituting Equations \ref{eq:est1} and \ref{eq:est2} into Equation \ref{eq:a2F_core} we propose an estimator for $\alpha^2\text{F}$, denoted as $\eta$, 
\begin{equation}
\eta = \left( \sum_a K_a\sqrt{\frac{M_0}{M_a}} \right)^2 \left( \sum_a K_a \right)^{-1} \left( \sum_s \frac{M_0}{M_s} \right)^{-1/2},
\label{eq:a2F_estimate}
\end{equation}
which can, considering  Equation \ref{eq:LC}, be used to predict {\tc} in the case of strong coupling. 

To determine the predictive power of the estimator $\eta$, we plot its value for a set of conventional superconductors versus \tc\ in  Figure~\ref{fig:E}. These results were obtained by either setting $\varepsilon_1$ to 0.5~eV, or to the energy of the bottom of the conduction band (if it was within 0.5~eV of $\varepsilon_\mathrm{F}$). This dataset included a number of elemental metals namely Nb, Li-fcc (35~GPa), Li-bcc, Cu, Ir, V, Tc, Sn, Al, and Pb (green squares)~\cite{kawamura2020benchmark}; the hydrides listed in Figure~\ref{fig:D} (orange triangles); and a set of ambient pressure conventional superconductors that contain covalent bonds consisting of  MgB$_2$~\cite{nagamatsu2001superconductivity}, NbN~\cite{shoji1992superconducting}, Nb$_3$Ge~\cite{gavaler1973superconductivity}, KPbB$_6$C$_6$~\cite{geng2023conventional}, SrB$_6$C$_6$~\cite{geng2023conventional}, NaC$_4$~\cite{PhysRevB.110.174508}, and YB$_6$~\cite{PhysRevB.73.024512} (blue circles). The dotted black line represents the best fit of the form
\begin{equation}
T_\mathrm{c}= \frac{a}{e^{b/\eta}-1},
\end{equation}
constructed by ensuring the exponential decay for small values of $\eta$, and linear behavior for high values of $\eta$, with $a = 13.1587$ and  $b = 0.00057578$, resulting in a coefficient of determination of $R^2=0.957$, and root mean square error of 20~K. The prefactor obtained here, though universal across all conventional superconductors, is expected to be sensitive to the choice of the pseudopotential and exchange correlation functional used. The \tc\ data originated from a variety of sources, including both theory and experiment, and therefore a perfect correlation is not to be expected. 
A subsequent set of experimentally synthesized superconductors consisting of Nb$_3$Al~\cite{naumov2015chemical}, V$_3$Si~\cite{ebihara2026high}, V$_3$Ge~\cite{ahmed2022superconducting}, TiN~\cite{krockenberger2012coherent},VN~\cite{luo1984superconductivity}, LuB$_{12}$~\cite{bolotina2018lower} and ZrB$_{12}$~\cite{bolotina2023low}, was chosen as testset for the validation of the fit, resulting in a root mean square error of 12~K.

As anticipated, and based upon the discussion in the previous sections, the model proposed here is able to provide a sensible value of {\tc} for most of the conventional superconductors we considered, with only V and AlH$_3$ as outliers. For vanadium, standard DFT calculations result in a {\tc} of $\sim$22~K, while spin-orbit and spin-fluctuations play an important role in bringing this prediction close to its experimental value of 6 K~\cite{kawamura2020benchmark}. Because these effects are not included in our analysis of the electronic structure, the $\eta$ parameter we obtain yields a \tc\ that is closer to the PBE value. For AlH$_3$ standard DFT calculations using the PBE exchange-correlation functional result in \tc s that are about one order of magnitude higher than those obtained including anharmonicity, and are closer to those predicted by our $\eta$ parameter. It should however be noted that more advanced DFT calculations predict the opening of a gap in AlH$_3$ at 150~GPa~\cite{shi2012electron}, and  experimentally no superconductivity was found above 4~K in AlH$_3$ at this pressure~\cite{PhysRevLett.100.045504}.

Because $\eta$ is quick to compute, we imagine it will become useful as an intermediate physical quantity to screen for superconductivity, or as a secondary objective to be used in Pareto-type CSP searches. Beyond its predictive capability, the descriptor $\eta$ (Equation \ref{eq:a2F_estimate}) provides a physical perspective on the factors that fundamentally constrain the superconducting transition temperature. In the present formulation, {\tc} is solely related to the strength of the interatomic interactions associated with the electronic states near the Fermi energy, and the magnitude of the atomic masses. The strength of the interatomic interactions is, in turn, governed by the chemistry of the constituent atoms, which therefore limits the maximum achievable {\tc} for a compound that contains a particular set of bonds. Pressure, on the other hand, provides a mechanism for modifying these interactions by reducing some interatomic distances thereby enhancing their orbital overlap and hardening the phonon modes with which they are associated, weakening other interactions and softening phonons, and finally by bringing certain electronic states to the Fermi level. Pressure can therefore increase both the characteristic phonon energy scale and the e--ph interaction strength. From this perspective, the pressure dependence of the proposed descriptor provides information not only on the expected variation of $T_c$, but also on the extent to which the underlying interactions can be engineered.

It has recently been shown that the contribution arising from the potential variation of the e-ph matrix elements plays an important role in suppressing the magnitude of the orbital overlap and restoring the local e-ph properties \cite{novoa2025real}. Such contributions can in general not be neglected, and it is possible that further refinements of our model that take into account this potential variation could further improve \tc\ prediction.

\section{Conclusions.}

This perspective investigates the relationship between chemical bonding and the physics governing conventional superconductivity. The magnitude of the superconducting critical temperature, \tc , is governed by the total integral of the Eliashberg spectral function, which we show is related to the magnitude of the local bonding strength between atoms integrated in a window around the Fermi energy. Previous formulations of this idea were based on heuristic application of the electron localization function as a means to capture the electron-phonon coupling in hydrides. 

Here, we justify why chemical-bonding based descriptors can be used as intermediate physical quantities for the prediction of \tc , and extend this methodology to all conventional superconductors. Specifically, we develop a descriptor that is based upon the integrated crystal orbital Hamilton population (COHP), which is chemically intuitive and computationally efficient, and therefore extremely useful for high throughput or crystal structure prediction searches of superconducting systems. It remains to be seen if this methodology can be extended to superconductors where the pairing mechanism is unclear. As a matter of fact, similar approaches for the description of  unconventional superconductors based on the real space overlap of orbitals have been proposed~\cite{bassen2025real}, but they are not yet capable of providing concrete \tc\ estimates.

In conclusion, the method presented here constitutes a powerful and versatile tool for the investigation of superconductivity in a conventional setting. Its accuracy, efficiency, and chemical interpretation make it suitable for practical applications, with the possibility that developments will broaden its predictive capabilities and enable its application to a wider range of superconducting materials.

\section{Methods}
DFT calculations were performed with the Vienna \emph{ab-initio} Simulation Package (VASP) version 6.4.2 \cite{Kresse:1993a, Kresse:1999a}. Projector Augmented-Wave (PAW) potentials~\cite{radevych2025rigid} were employed together with the Perdew-Burke-Ernzerhof (PBE)~\cite{GGA-PBE} exchange-correlation functional. The employed PAW electronic valence configurations were 
Al 3s$^{2}$3p$^{1}$,
B 2s$^{2}$2p$^{1}$,
C 2s$^{2}$2p$^{2}$,
Ca 3s$^{2}$3p$^{6}$4s$^{2}$,
Cu 3d$^{10}$4s$^{1}$,
Ge 4s$^{2}$4p$^{2}$,
H 1s$^{1}$2p$^{0}$,
Ir 5s$^{1}$6d$^{8}$,
K 3s$^{2}$3p$^{6}$4s$^{1}$,
La 5s$^{2}$6s$^{2}$5p$^{6}$5d$^{1}$,
Lu 4f$^{14}$5s$^{2}$6s$^{2}$5p$^{6}$5d$^{1}$
Mg 2p$^{6}$3s$^{2}$,
Mo 4p$^{6}$5s$^{0}$4d$^{0}$,
N 2s$^{2}$2p$^{3}$,
Na 3s$^{1}$3p$^{0}$,
S 3s$^{2}$3p$^{4}$,
Sc 3p$^{6}$4s$^{1}$3d$^{0}$,
 Sn 5s$^{2}$5p$^{2}$,
Sr 4s$^{2}$4p$^{6}$5s$^{2}$,
Tc 4p$^{6}$5s$^{2}$,
Ti 3d$^{3}$4s$^{2}$
V 3p$^{6}$3d$^{4}$4s$^{1}$, 
Y 4s$^{2}$4p$^{6}$5s$^{2}$4d$^{2}$, and
Zr 4s$^{2}$5s$^{2}$4p$^{6}$4d$^{3}$
Converged parameters for the calculations considered an energy cutoff of 600~eV, a $\mathbf{k}$-point grid density of $2\pi \times 0.025$ \AA\ $^{-1}$, and a first-order Methfessel-Paxton \cite{Methfessel} smearing of 0.2~eV.
The energy resolved bond projected COHP was obtained through the Local Orbital Basis Suite Towards Electronic-Structure Reconstruction (LOBSTER) version 5.5.1~\cite{maintz2016lobster} using the tetrahedron method for the mesh construction and the \emph{vasp2015} basis/projector set. The obtained charge spilling was found to be greater than 4\% but lower than 10\% for Ir, LaH$_{10}$, Li-bcc, MgIrH$_6$, Pb, Sn,V$_3$Ge,V$_3$Si and V,  but less than 4\% for the rest of the systems studied.
\section*{Acknowledgements}
We thank Trevor Higgins and Chris Walsh for their insightful suggestions and benchmarking of the pCOHP descriptor. We further acknowledge the National Science Foundation (Grant No.\ DMR-2136038) and the U.S. Department of
Energy, Office of Science, Fusion Energy Sciences, funding the award entitled ``High Energy Density Quantum Matter,'' Award
No.\ DE-SC0020340 for financial support. Calculations were performed at the Center for Computational Research at SUNY Buffalo (https://hdl.handle.net/10477/79221).


\bibliography{bibliography,zurek}

@article{Onnes,
  title={The Resistance of Pure Mercury at Helium Temperatures},
  author={Onnes, K. H. T.},
  journal={Commun. Phys. Lab Univ Leiden B},
  volume={12},
  number={1},
  year={1911},
}

@article{racioppi2023electride,
	title={On the Electride Nature of Na-hP4},
	author={Racioppi, Stefano and Storm, Christian V and McMahon, Malcolm I and Zurek, Eva},
	journal={Angew. Chem.},
	volume={62},
	number={48},
	pages={e202310802},
	year={2023},
	publisher={Wiley Online Library}
}

@inproceedings{shi2012electron,
	title={Electron band structure of the high pressure cubic phase of AlH3},
	author={Shi, Hongliang and Zarifi, Niliffar and Yim, Wai-Leung and Tse, JS},
	booktitle={J. Phys: Conf. Ser.},
	volume={377},
	number={1},
	pages={012093},
	year={2012}
}

@article{PhysRevLett.100.045504,
	title = {Pressure-Induced Hydrogen-Dominant Metallic State in Aluminum Hydride},
	author = {Goncharenko, Igor and Eremets, M. I. and Hanfland, M. and Tse, J. S. and Amboage, M. and Yao, Y. and Trojan, I. A.},
	journal = {Phys. Rev. Lett.},
	volume = {100},
	issue = {4},
	pages = {045504},
	numpages = {4},
	year = {2008},
	month = {Jan},
	publisher = {American Physical Society},
	doi = {10.1103/PhysRevLett.100.045504},
	url = {https://link.aps.org/doi/10.1103/PhysRevLett.100.045504}
}

@article{PhysRev.78.635,
  title = {Superconductivity of Lead},
  author = {Boorse, H. A. and Cook, D. B. and Zemansky, M. W.},
  journal = {Phys. Rev.},
  volume = {78},
  issue = {5},
  pages = {635--636},
  numpages = {0},
  year = {1950},
  month = {Jun},
  publisher = {American Physical Society},
  doi = {10.1103/PhysRev.78.635},
  url = {https://link.aps.org/doi/10.1103/PhysRev.78.635}
}

@article{Kresse:1993a,
	title        = {\textit{Ab initio} Molecular Dynamics for Liquid Metals},
	author       = {Kresse, G. and    Hafner, J.},
	year         = 1993,
	journal      = {Phys. Rev. B.},
	volume       = 47,
	pages        = {558--561}
}

@article{Kresse:1999a,
	title        = {From Ultrasoft Pseudopotentials to the Projector Augmented--Wave Method},
	author       = {Kresse, G. and    Joubert, D.},
	year         = 1999,
	journal      = {Phys. Rev. B.},
	volume       = 59,
	pages        = {1758--1775}
}

@ARTICLE{GGA-PBE,
	title={Generalized gradient approximation made simple},
	author={Perdew, J. P. and Burke, K. and    Ernzerhof, M.},
	journal={Phys. Rev. Lett.},
	volume={77},
	number={18},
	pages={3865},
	year={1996},
	publisher={APS}
}

@ARTICLE{Methfessel,
	title={High-precision sampling for Brillouin-zone integration in metals},
	author={Methfessel, M. and    Paxton, A.},
	journal={Phys. Rev. B},
	volume={40},
	number={6},
	pages={3616},
	year={1989},
	publisher={APS}
}

@article{maintz2016lobster,
	title={LOBSTER: A tool to extract chemical bonding from plane-wave based DFT},
	author={Maintz, Stefan and Deringer, Volker L and Tchougr{\'e}eff, Andrei L and Dronskowski, Richard},
	journal={J. Comput. Chem.},
	volume={37},
	number={11},
	pages={1030--1035},
	year={2016},
	publisher={Wiley Online Library},
	note={Software version 5.5.1}
}

@article{wines2024data,
  title={Data-Driven Design of High Pressure Hydride Superconductors Using DFT and Deep Learning},
  author={Wines, Daniel and Choudhary, Kamal},
  journal={Mater. futures},
  volume={3},
  number={2},
  pages={025602},
  year={2024},
  publisher={IOP Publishing}
}

@article{PhysRevB.75.045110,
  title = {Construction of Wannier Functions from Localized Atomiclike Orbitals},
  author = {Solovyev, I. V. and Pchelkina, Z. V. and Anisimov, V. I.},
  journal = {Phys. Rev. B},
  volume = {75},
  issue = {4},
  pages = {045110},
  numpages = {10},
  year = {2007},
  month = {Jan},
  publisher = {American Physical Society},
  doi = {10.1103/PhysRevB.75.045110},
  url = {https://link.aps.org/doi/10.1103/PhysRevB.75.045110}
}

@article{ahmed2022superconducting,
	title={Superconducting properties of the A15 structure compound V$_3$Ge},
	author={Ahmed, Esteaque and Kobayashi, Kazu and Arakawa, Naoya and Okubo, Yuta and Sakai, Akito and Nakatsuji, Satoru and Iyo, Akira and Ebihara, Takao},
	journal={Phys. C: Supercond. Appl.},
	volume={602},
	pages={1354140},
	year={2022},
	publisher={Elsevier}
}

@article{krockenberger2012coherent,
	title={Coherent growth of superconducting TiN thin films by plasma enhanced molecular beam epitaxy},
	author={Krockenberger, Yoshiharu and Karimoto, Shin-ichi and Yamamoto, Hideki and Semba, Kouich},
	journal={J. Appl. Phys.},
	volume={112},
	number={8},
	year={2012},
	publisher={AIP Publishing}
}

@article{luo1984superconductivity,
	title={Superconductivity of VN under pressure},
	author={Luo, HL and Wolf, SA and Fuller, WW and Edelstein, AS and Huang, CY},
	journal={Phys. Rev. B},
	volume={29},
	number={3},
	pages={1443},
	year={1984},
	publisher={APS}
}

@article{bolotina2023low,
	title={Low temperature singularities of electron density in a two-gap superconductor ZrB$_{12}$},
	author={Bolotina, NB and Khrykina, ON and Azarevich, AN and Shitsevalova, N Yu and Filipov, VB and Gavrilkin, S Yu and Tsvetkov, A Yu and Gab{\'a}ni, S and Flachbart, K and Voronov, VV and others},
	journal={Solid State Sci.},
	volume={142},
	pages={107245},
	year={2023},
	publisher={Elsevier}
}

@article{bolotina2018lower,
	title={The lower symmetry electron-density distribution and the charge transport anisotropy in cubic dodecaboride LuB$_{12}$},
	author={Bolotina, Nadezhda B and Dudka, Alexander P and Khrykina, Olga N and Krasnorussky, Vladimir N and Shitsevalova, Natalya Yu and Filipov, Volodymyr B and Sluchanko, Nikolay E},
	journal={Jo. Phys. Condens. Matter},
	volume={30},
	number={26},
	pages={265402},
	year={2018},
	publisher={IOP Publishing}
}

@article{ebihara2026high,
	title={High-Field Specific Heat and Fundamental Superconducting Properties of the Single-Phase Type-II Superconductor V$_3$Si},
	author={Ebihara, Takao and Kobayashi, Kazu and Asakura, Mihiro and Sakai, Akito and Nakatsuji, Satoru and Iyo, Akira and Kohama, Yoshimitsu},
	journal={J.Phys. Soc. of Jpn.},
	volume={95},
	number={10},
	pages={104707},
	year={2026},
	publisher={The Physical Society of Japan}
}

@article{naumov2015chemical,
  title={Chemical Bonding in Hydrogen and Lithium Under Pressure},
  author={Naumov, Ivan I and Hemley, Russell J and Hoffmann, Roald and Ashcroft, NW},
  journal={J. Chem. Phys.},
  volume={143},
  number={6},
  year={2015},
  publisher={AIP Publishing}
}

@article{RevModPhys.32.300,
  title = {Canonical Configurational Interaction Procedure},
  author = {Foster, J. M. and Boys, S. F.},
  journal = {Rev. Mod. Phys.},
  volume = {32},
  issue = {2},
  pages = {300--302},
  numpages = {0},
  year = {1960},
  month = {Apr},
  publisher = {American Physical Society},
  doi = {10.1103/RevModPhys.32.300},
  url = {https://link.aps.org/doi/10.1103/RevModPhys.32.300}
}

@article{da2025machine,
  title={Machine-Learning Accelerated Prediction of Two-Dimensional Conventional Superconductors},
  author={da Silva, Thalis HB and Cavignac, Th{\'e}o and Cerqueira, Tiago FT and Wang, Hai-Chen and Marques, Miguel AL},
  journal={Mater. Horiz.},
  volume={12},
  number={10},
  pages={3408--3419},
  year={2025},
  publisher={Royal Society of Chemistry}
}

@article{cerqueira2024sampling,
  title={Sampling the Materials Space for Conventional Superconducting Compounds},
  author={Cerqueira, Tiago FT and Sanna, Antonio and Marques, Miguel AL},
  journal={Adv, Mater.},
  volume={36},
  number={1},
  pages={2307085},
  year={2024},
  publisher={Wiley Online Library}
}

@article{wines2023inverse,
  title={Inverse Design of Next-Generation Superconductors Using Data-Driven Deep Generative Models},
  author={Wines, Daniel and Xie, Tian and Choudhary, Kamal},
  journal={J. Phys. Chem. Lett.},
  volume={14},
  number={29},
  pages={6630--6638},
  year={2023},
  publisher={ACS Publications}
}

@article{wines2023high,
  title={High-Throughput DFT-Based Discovery of Next Generation Two-Dimensional (2D) Superconductors},
  author={Wines, Daniel and Choudhary, Kamal and Biacchi, Adam J and Garrity, Kevin F and Tavazza, Francesca},
  journal={Nano lett.},
  volume={23},
  number={3},
  pages={969--978},
  year={2023},
  publisher={ACS Publications}
}

@article{prakash2025guided,
  title={Guided Diffusion for the Discovery of New Superconductors},
  author={Prakash, Pawan and Gibson, Jason B and Li, Zhongwei and Di Gianluca, Gabriele and Esquivel, Juan and Fuemmeler, Eric and Geisler, Benjamin and Kim, Jung Soo and Roitberg, Adrian and Tadmor, Ellad B and others},
  journal={arXiv preprint arXiv:2509.25186},
  year={2025}
}

@article{Geisler:2026a,
  title={Towards the Discovery of High Critical Magnetic Field Superconductors},
  author={Geisler, B. and Dee, P. M. and Hamlin, J. J. and Stewart, G. R. and Hennig, R. G. and Hirschfeld, P. J.},
  journal={npj Comput. Mater.},
  year={2026},
  url={https://doi.org/10.1038/s41524-026-02266-9},
  note={In press.}
}

@article{gibson2503developing,
  title={Developing a complete {{AI}}-accelerated workflow for superconductor discovery},
  author={Gibson, J. B. and Hire, A. C. and Prakash, P. and Dee, P. M. and Geisler, B. and Kim, J. S. and Li, Z. and Hamlin, J. J. and Stewart, G. R. and Hirschfeld, P. J. and Hennig, R. G.},
  journal={npj Comput. Mater.},
  volume={12},
  number={1},
  pages={95},
  year={2026},
  publisher={Nature Publishing Group UK London}
}

@article{choudhary2022designing,
  title={Designing High-T$_c$ Superconductors with BCS-Inspired Screening, Density Functional Theory, and Deep-Learning},
  author={Choudhary, Kamal and Garrity, Kevin},
  journal={npj Comput. Mater.},
  volume={8},
  number={1},
  pages={244},
  year={2022},
  publisher={Nature Publishing Group UK London}
}

@article{PhysRev.95.1435,
  title = {Superconductivity of $\mathrm{Nb}_{3}$Sn},
  author = {Matthias, B. T. and Geballe, T. H. and Geller, S. and Corenzwit, E.},
  journal = {Phys. Rev.},
  volume = {95},
  issue = {6},
  pages = {1435--1435},
  numpages = {0},
  year = {1954},
  month = {Sep},
  publisher = {American Physical Society},
  doi = {10.1103/PhysRev.95.1435},
  url = {https://link.aps.org/doi/10.1103/PhysRev.95.1435}
}

@ARTICLE {Livas,
  title = { A Perspective on Conventional High-Temperature Superconductors at High Pressure: Methods and Materials},
  author={Flores-Livas, Jos{\'e} A and Boeri, Lilia and Sanna, Antonio and Profeta, Gianni and Arita, Ryotaro and Eremets, Mikhail},
  journal={Phys. Rep.},
  volume={856},
  pages={1--78},
  year={2020},
  publisher={Elsevier}
}

@article{gavaler1973superconductivity,
  title={Superconductivity in Nb-Ge Films Above 22 K},
  author={Gavaler, JR},
  journal={Appl. Phys. Lett.},
  volume={23},
  number={8},
  pages={480--482},
  year={1973}
}

@article{shoji1992superconducting,
  title={Superconducting Properties and Normal-State Resistivity of Single-Crystal NbN Films Prepared by a Reactive rf-Magnetron Sputtering Method},
  author={Shoji, A and Kiryu, S and Kohjiro, S},
  journal={Appl. phys. lett.},
  volume={60},
  number={13},
  pages={1624--1626},
  year={1992},
  publisher={American Institute of Physics}
}

@article{nagamatsu2001superconductivity,
  title={Superconductivity at 39 K in Magnesium Diboride.},
  author={Nagamatsu, Jun and Nakagawa, Norimasa and Muranaka, Takahiro and Zenitani, Yuji and Akimitsu, Jun},
  journal={nature},
  volume={410},
  number={6824},
  pages={63},
  year={2001}
}

@Article{Drozdov:2015a,
  author    = {Drozdov, A. P. and Eremets, M. I. and Troyan, I. A. and Ksenofontov, V and Shylin, S. I},
  title = { Conventional Superconductivity at 203 Kelvin at High Pressures in the Sulfur Hydride System},
  journal   = {Nature},
  year      = {2015},
  volume    = {525},
  pages     = {73-76},
  publisher = {Nature Publishing Group},
}

@article{LaH10_Somayazulu,
  title = {Evidence for Superconductivity Above 260 K in Lanthanum Superhydride at Megabar Pressures},
  author = {Somayazulu, Maddury and Ahart, Muhtar and Mishra, Ajay K. and Geballe, Zachary M. and Baldini, Maria and Meng, Yue and Struzhkin, Viktor V. and Hemley, Russell J.},
  journal = {Phys. Rev. Lett.},
  volume = {122},
  issue = {2},
  pages = {027001},
  numpages = {6},
  year = {2019},
  month = {Jan},
  publisher = {American Physical Society},
  doi = {10.1103/PhysRevLett.122.027001},
  url = {https://link.aps.org/doi/10.1103/PhysRevLett.122.027001}
}

@article{drozdov2019superconductivity,
  title={Superconductivity at 250 K in Lanthanum Hydride Under High Pressures},
  author={Drozdov, AP and Kong, PP and Minkov, VS and Besedin, SP and Kuzovnikov, MA and Mozaffari, S and Balicas, L and Balakirev, Fedor Fedorovich and Graf, DE and Prakapenka, VB and others},
  journal={Nature},
  volume={569},
  number={7757},
  pages={528--531},
  year={2019},
  publisher={Nature Publishing Group UK London}
}

@article{PhysRev.108.1175,
  title = {Theory of Superconductivity},
  author = {Bardeen, J. and Cooper, L. N. and Schrieffer, J. R.},
  journal = {Phys. Rev.},
  volume = {108},
  issue = {5},
  pages = {1175--1204},
  numpages = {0},
  year = {1957},
  month = {Dec},
  publisher = {American Physical Society},
  doi = {10.1103/PhysRev.108.1175},
  url = {https://link.aps.org/doi/10.1103/PhysRev.108.1175}
}

@article{migdal1958interaction,
  title={Interaction Between Electrons and Lattice Vibrations in a Normal Metal},
  author={Migdal, AB},
  journal={Sov. Phys. JETP},
  volume={7},
  number={6},
  pages={996--1001},
  year={1958}
}

@article{eliashberg1960interactions,
  title={Interactions Between Electrons and Lattice Vibrations in a Superconductor},
  author={Eliashberg, GM},
  journal={Sov. Phys. JETP},
  volume={11},
  number={3},
  pages={696--702},
  year={1960}
}

@article{eliashberg1961temperature,
  title={Temperature Green's Function for Electrons in a Superconductor},
  author={Eliashberg, GM},
  journal={Sov. Phys. JETP},
  volume={12},
  number={5},
  pages={1000--1002},
  year={1961}
}

@article{annurev:/content/journals/10.1146/annurev-conmatphys-070909-104117,
   author = "Taillefer, Louis",
   title = "Scattering and Pairing in Cuprate Superconductors", 
   journal= "Annu. Rev. Condens. Matter Phys.",
   year = "2010",
   volume = "1",
   pages = "51-70",
   doi = "https://doi.org/10.1146/annurev-conmatphys-070909-104117",
   url = "https://www.annualreviews.org/content/journals/10.1146/annurev-conmatphys-070909-104117",
   publisher = "Annual Reviews",
   issn = "1947-5462",
   type = "Journal Article",

  }

@inproceedings{negi2024cuprate,
  title={Cuprate Superconductors: An Overview},
  author={Negi, Minal and Singh, Amit Raj and Vikram, Kunwar and Roy, Soumendra K and Singh, Anjani Kumar},
  booktitle={International Conference on Recent Advances in Functional Materials},
  pages={43--70},
  year={2024},
  organization={Springer}
}

@article{RevModPhys.83.1589,
  title = {Superconductivity in Iron Compounds},
  author = {Stewart, G. R.},
  journal = {Rev. Mod. Phys.},
  volume = {83},
  issue = {4},
  pages = {1589--1652},
  numpages = {0},
  year = {2011},
  month = {Dec},
  publisher = {American Physical Society},
  doi = {10.1103/RevModPhys.83.1589},
  url = {https://link.aps.org/doi/10.1103/RevModPhys.83.1589}
}

@article{Troyan,
  title = { Anomalous High-Temperature Superconductivity in YH$_6$},
  author={Troyan, Ivan A. and Semenok, Dmitrii V. and Kvashnin, Alexander G. and Sadakov, Andrey V. and Sobolevskiy, Oleg A. and Pudalov, Vladimir M. and Ivanova, Anna G. and Prakapenka, Vitali B. and Greenberg, Eran and Gavriliuk, Alexander G. and Lyubutin, Igor S. and Struzhkin, Viktor V. and Bergara, Aitor and Errea, Ion and Bianco, Raffaello and Calandra, Matteo and Mauri, Francesco and Monacelli, Lorenzo and Akashi, Ryosuke and Oganov, Artem R.},
  journal={Adv. Mater.},
  volume={33},
  number={15},
  pages={2006832},
  year={2021},
  publisher={Wiley Online Library}
}

@ARTICLE{atomicH,
  title = { Anharmonic Effects in Atomic Hydrogen: Superconductivity and Lattice Dynamical Stability},
  author={Borinaga, Miguel and Errea, Ion and Calandra, Matteo and Mauri, Francesco and Bergara, Aitor},
  journal={Phys. Rev. B},
  volume={93},
  number={17},
  pages={174308},
  year={2016},
  publisher={APS}
}

@article{errea2020quantum,
  title = { Quantum Crystal Structure in the 250-Kelvin Superconducting Lanthanum Hydride},
  author={Errea, Ion and Belli, Francesco and Monacelli, Lorenzo and Sanna, Antonio and Koretsune, Takashi and Tadano, Terumasa and Bianco, Raffaello and Calandra, Matteo and Arita, Ryotaro and Mauri, Francesco and Flores-Livas, Jos{\'{e}} A.},
  journal={Nature},
  volume={578},
  number={7793},
  pages={66--69},
  year={2020},
  publisher={Nature Publishing Group UK London}
}

@article{kitatani2020nickelate,
  title={Nickelate Superconductors-a Renaissance of the One-Band Hubbard Model},
  author={Kitatani, Motoharu and Si, Liang and Janson, Oleg and Arita, Ryotaro and Zhong, Zhicheng and Held, Karsten},
  journal={npj Quantum Mater.},
  volume={5},
  number={1},
  pages={59},
  year={2020},
  publisher={Nature Publishing Group UK London}
}

@article{brison2000heavy,
  title={Heavy Fermion Superconductivity},
  author={Brison, Jean-Pascal and Gl{\'e}mot, Lo{\i}\"c and Suderow, Hermann and Huxley, Andrew and Kambe, Shinsaku and Flouquet, Jacques},
  journal={Phys. Rev. B Condens. Matter},
  volume={280},
  number={1-4},
  pages={165--171},
  year={2000},
  publisher={Elsevier}
}

@article{ferreira2024ab,
  title={Ab Initio Modeling of Superconducting Alloys},
  author={Ferreira, Pedro N and Lucrezi, Roman and Guilhon, Ivan and Marques, Marcelo and Teles, Lara K and Heil, Christoph and Eleno, Luiz TF},
  journal={Mater. Today Phys.},
  volume={48},
  pages={101547},
  year={2024},
  publisher={Elsevier}
}

@article{lucrezi2024full,
  title={Full-bandwidth anisotropic Migdal-Eliashberg theory and its application to superhydrides},
  author={Lucrezi, Roman and Ferreira, Pedro P and Hajinazar, Samad and Mori, Hitoshi and Paudyal, Hari and Margine, Elena R and Heil, Christoph},
  journal={Commun. Phys.},
  volume={7},
  number={1},
  pages={33},
  year={2024},
  publisher={Nature Publishing Group UK London}
}

@article{PhysRevLett.111.177002,
  title = {First-Principles Theory of Anharmonicity and the Inverse Isotope Effect in Superconducting Palladium-Hydride Compounds},
  author = {Errea, Ion and Calandra, Matteo and Mauri, Francesco},
  journal = {Phys. Rev. Lett.},
  volume = {111},
  issue = {17},
  pages = {177002},
  numpages = {5},
  year = {2013},
  month = {Oct},
  publisher = {American Physical Society},
  doi = {10.1103/PhysRevLett.111.177002},
  url = {https://link.aps.org/doi/10.1103/PhysRevLett.111.177002}
}

@article{scanlan2004superconducting,
  title={Superconducting Materials for Large Scale Applications},
  author={Scanlan, Ronald M and Malozemoff, Alexis P and Larbalestier, David C},
  journal={Proc. IEEE},
  volume={92},
  number={10},
  pages={1639--1654},
  year={2004},
  publisher={IEEE}
}

@article{yao2021superconducting,
  title={Superconducting Materials: Challenges and Opportunities for Large-Scale Applications},
  author={Yao, Chao and Ma, Yanwei},
  journal={Iscience},
  volume={24},
  number={6},
  year={2021},
  publisher={Elsevier}
}

@article{PhysRevB.87.024505,
  title = {Anisotropic Migdal-Eliashberg theory using Wannier functions},
  author = {Margine, E. R. and Giustino, F.},
  journal = {Phys. Rev. B},
  volume = {87},
  issue = {2},
  pages = {024505},
  numpages = {12},
  year = {2013},
  month = {Jan},
  publisher = {American Physical Society},
  doi = {10.1103/PhysRevB.87.024505},
  url = {https://link.aps.org/doi/10.1103/PhysRevB.87.024505}
}

@article{PhysRevB.76.165108,
  title = {Electron-Phonon Interaction Using Wannier Functions},
  author = {Giustino, Feliciano and Cohen, Marvin L. and Louie, Steven G.},
  journal = {Phys. Rev. B},
  volume = {76},
  issue = {16},
  pages = {165108},
  numpages = {19},
  year = {2007},
  month = {Oct},
  publisher = {American Physical Society},
  doi = {10.1103/PhysRevB.76.165108},
  url = {https://link.aps.org/doi/10.1103/PhysRevB.76.165108}
}

@article{bianco2023non,
  title={Non-Perturbative Theory of the Electron-Phonon Coupling and its First-Principles Implementation},
  author={Bianco, Raffaello and Errea, Ion},
  journal={arXiv preprint arXiv:2303.02621},
  year={2023}
}

@article{belli2025efficient,
  title={Efficient Modelling of Anharmonicity and Quantum Effects in PdCuH$_2$ with Machine Learning Potentials},
  author={Belli, Francesco and Zurek, Eva},
  journal={npj Comput. Mater.},
  volume={11},
  number={1},
  pages={87},
  year={2025},
  publisher={Nature Publishing Group UK London}
}

@article{belli2025chemical,
  title={A Chemical Bonding Based Descriptor for Predicting the Impact of Quantum Nuclear and Anharmonic Effects on Hydrogen-Based Superconductors},
  author={Belli, Francesco and Zurek, Eva and Errea, Ion},
  journal={arXiv preprint arXiv:2501.14420},
  year={2025}
}

@ARTICLE{H_mol,
  title = { Anharmonic Enhancement of Superconductivity in Metallic Molecular $Cmca\bar{4}$ Hydrogen at High Pressure: A First-Principles Study},
  author={Borinaga, Miguel and Riego, P and Leonardo, Aritz and Calandra, Matteo and Mauri, Francesco and Bergara, Aitor and Errea, Ion},
  journal={J. Phys. Condens. Matter},
  volume={28},
  number={49},
  pages={494001},
  year={2016},
  publisher={IOP Publishing}
}

@article{errea2015high,
  title = { High-Pressure Hydrogen Sulfide From First Principles: A Strongly Anharmonic Phonon-Mediated Superconductor},
  author={Errea, Ion and Calandra, Matteo and Pickard, Chris J and Nelson, Joseph and Needs, Richard J and Li, Yinwei and Liu, Hanyu and Zhang, Yunwei and Ma, Yanming and Mauri, Francesco},
  journal={Phys. Rev. Lett.},
  volume={114},
  number={15},
  pages={157004},
  year={2015},
  publisher={APS}
}

@article{hou2021strong,
  title = { Strong Anharmonic and Quantum Effects in Pm$\bar{3}$n AlH$_{3}$ Under High Pressure: A First-Principles Study},
  author={Hou, Pugeng and Belli, Francesco and Bianco, Raffaello and Errea, Ion},
  journal={Phys. Rev. B},
  volume={103},
  number={13},
  pages={134305 (1-8)},
  year={2021},
  publisher={APS}
}

@article{Hou3,
  title = { Quantum Anharmonic Enhancement of Superconductivity in $P6_3/mmc$ ScH$_6$ at High Pressures: A First-Principles Study},
  author={Hou, Pugeng and Belli, Francesco and Bianco, Raffaello and Errea, Ion},
  journal={J. Appl. Phys.},
  volume={130},
  number={17},
  pages={175902},
  year={2021},
  publisher={AIP Publishing LLC}
}

@article{pickard2011ab,
  title={Ab Initio Random Structure Searching},
  author={Pickard, Chris J and Needs, RJ},
  journal={J. Phys.: Condens. Mat.},
  volume={23},
  number={5},
  pages={053201},
  year={2011},
  publisher={IOP Publishing}
}

@article{mishra2025electron,
	title={Electron-phonon vertex correction effect in superconducting H$_3$S},
	author={Mishra, Shashi B and Mori, Hitoshi and Margine, Elena R},
	journal={npj Computational Materials},
	volume={11},
	number={1},
	pages={342},
	year={2025},
	publisher={Nature Publishing Group UK London}
}

@article{pickett2025compressed,
	title={Why Compressed Hydrides Are Near-Room-Temperature Superconductors},
	author={Pickett, Warren E},
	journal={Ann. Phys. (Berl.)},
	volume={538},
	number={1},
	pages={e00362},
	year={2026},
	publisher={Wiley Online Library}
}

@article{wang2012calypso,
  title={CALYPSO: A Method for Crystal Structure Prediction},
  author={Wang, Yanchao and Lv, Jian and Zhu, Li and Ma, Yanming},
  journal={Comput. Phys. Commun.},
  volume={183},
  number={10},
  pages={2063-2070},
  year={2012},
  publisher={Elsevier}
}

@article{oganov2006crystal,
  title={Crystal Structure Prediction Using Ab Initio Evolutionary Techniques: Principles and Applications},
  author={Oganov, Artem R and Glass, Colin W},
  journal={J. Chem. Phys.},
  volume={124},
  number={24},
  year={2006},
  publisher={AIP Publishing}
}

@article{lonie2011xtalopt,
  title={XtalOpt: An Open-Source Evolutionary Algorithm for Crystal Structure Prediction},
  author={Lonie, David C and Zurek, Eva},
  journal={Comput. Phys. Commun.},
  volume={182},
  number={2},
  pages={372-387},
  year={2011},
  publisher={Elsevier}
}

@article{wang2024mg2irh6,
  title={Why Mg$_2$IrH$_6$ is Predicted to be a High-Temperature Superconductor, but Ca$_2$IrH$_6$ is not},
  author={Wang, Xiaoyu and Pickett, Warren E and Hutcheon, Michael and Prasankumar, Rohit P and Zurek, Eva},
  journal={Angew. Chem.},
  volume={136},
  number={52},
  pages={e202412687},
  year={2024},
  publisher={Wiley Online Library}
}

@article{hajinazar2025xtalopt,
  title={XtalOpt Version 14: Variable-Composition Crystal Structure Search for Functional Materials Through Pareto Optimization},
  author={Hajinazar, Samad and Zurek, Eva},
  journal={arXiv preprint arXiv:2506.17246},
  year={2025}
}

@article{PhysRevB.106.L180501,
  title = {Why Mercury is a Superconductor},
  author = {Tresca, Cesare and Profeta, Gianni and Marini, Giovanni and Bachelet, Giovanni B. and Sanna, Antonio and Calandra, Matteo and Boeri, Lilia},
  journal = {Phys. Rev. B},
  volume = {106},
  issue = {18},
  pages = {L180501},
  numpages = {6},
  year = {2022},
  month = {Nov},
  publisher = {American Physical Society},
  doi = {10.1103/PhysRevB.106.L180501},
  url = {https://link.aps.org/doi/10.1103/PhysRevB.106.L180501}
}

@article{salzbrenner2023developments,
  title={Developments and Further Applications of Ephemeral Data Derived Potentials},
  author={Salzbrenner, Pascal T and Joo, Se Hun and Conway, Lewis J and Cooke, Peter IC and Zhu, Bonan and Matraszek, Milosz P and Witt, William C and Pickard, Chris J},
  journal={J. Chem. Phys.},
  volume={159},
  number={14},
  year={2023},
  publisher={AIP Publishing}
}

@article{pickard2022ephemeral,
  title={Ephemeral Data Derived Potentials for Random Structure Search},
  author={Pickard, Chris J},
  journal={Phys. Rev. B},
  volume={106},
  number={1},
  pages={014102},
  year={2022},
  publisher={APS}
}

@article{luo2024deep,
  title={Deep Learning Generative Model for Crystal Structure Prediction},
  author={Luo, Xiaoshan and Wang, Zhenyu and Gao, Pengyue and Lv, Jian and Wang, Yanchao and Chen, Changfeng and Ma, Yanming},
  journal={npj Comput. Mater.},
  volume={10},
  number={1},
  pages={254},
  year={2024},
  publisher={Nature Publishing Group UK London}
}

@article{lucrezi2022silico,
  title={In-Silico Synthesis of Lowest-Pressure high-T$_\mathrm{c}$ Ternary Superhydrides},
  author={Lucrezi, Roman and Di Cataldo, Simone and von der Linden, Wolfgang and Boeri, Lilia and Heil, Christoph},
  journal={npj Comput. Mater.},
  volume={8},
  number={1},
  pages={119},
  year={2022},
  publisher={Nature Publishing Group UK London}
}

@article{cataldo2022,
  title={First-principles search of hot superconductivity in La-X-H Ternary Hydrides},
  author={Di Cataldo, Simone and von der Linden, Wolfgang and Boeri, Lilia},
  journal={npj Comput. Mater.},
  volume={8},
  pages={2},
  year={2022},
  publisher={Nature Publishing Group UK London}
}

@article{GENG2024101443,
title = {Superconductivity in CH$_4$ and BH$_4^-$ Containing Compounds Derived from the High-Pressure Superhydrides},
journal = {Mater. Today Phy.},
volume = {44},
pages = {101443},
year = {2024},
issn = {2542-5293},
doi = {https://doi.org/10.1016/j.mtphys.2024.101443},
url = {https://www.sciencedirect.com/science/article/pii/S2542529324001196},
author = {Nisha Geng and Katerina P. Hilleke and Francesco Belli and Pratik Kumar Das and Eva Zurek},

}

@article{gibson2025accelerating,
  title={Accelerating Superconductor Discovery Through Tempered Deep Learning of the Electron-Phonon Spectral Function},
  author={Gibson, Jason B and Hire, Ajinkya C and Dee, Philip M and Barrera, Oscar and Geisler, Benjamin and Hirschfeld, Peter J and Hennig, Richard G},
  journal={npj Comput. Mater.},
  volume={11},
  number={1},
  pages={7},
  year={2025},
  publisher={Nature Publishing Group UK London}
}

@article{MA2023101233,
title = {High-Throughput Calculation for Superconductivity of Sodalite-Like Clathrate Ternary Hydrides MXH$_12$ at High Pressure},
journal = {Mater. Today Phys.},
volume = {38},
pages = {101233},
year = {2023},
issn = {2542-5293},
doi = {https://doi.org/10.1016/j.mtphys.2023.101233},
url = {https://www.sciencedirect.com/science/article/pii/S2542529323002699},
author = {Tiancheng Ma and Zihan Zhang and Mingyang Du and Zihao Huo and Wenxuan Chen and Fubo Tian and Defang Duan and Tian Cui},
}

@article{kawamura2020benchmark,
  title={Benchmark of Density Functional Theory for Superconductors in Elemental Materials},
  author={Kawamura, Mitsuaki and Hizume, Yuma and Ozaki, Taisuke},
  journal={Phys. Rev. B},
  volume={101},
  number={13},
  pages={134511},
  year={2020},
  publisher={APS}
}

@article{novoa2025real,
  title = {Real-Space Understanding of Electron-Phonon Coupling in Superconducting Hydrides},
  author = {Novoa, Trinidad and Belli, Francesco and Bianco, Raffaello and Contreras-Garc{\'{\i}}a, Julia and Errea, Ion},
  journal = {Phys. Rev. B},
  volume = {114},
  issue = {15},
  pages = {154512},
  numpages = {12},
  year = {2026},
  month = {Sep},
  publisher = {American Physical Society},
  doi = {10.1103/6w1y-jwfm},
  url = {https://link.aps.org/doi/10.1103/6w1y-jwfm}
}

@article{PhysRevB.110.174515,
  title = {Molecularity: A Fast and Efficient Criterion for Probing Superconductivity},
  author = {di Mauro, Mat\'{\i}as E. and Bra\"{\i}da, Beno\^{\i}t and Errea, Ion and Novoa, Trinidad and Contreras-Garc\'{\i}a, Julia},
  journal = {Phys. Rev. B},
  volume = {110},
  issue = {17},
  pages = {174515},
  numpages = {9},
  year = {2024},
  month = {Nov},
  publisher = {American Physical Society},
  doi = {10.1103/PhysRevB.110.174515},
  url = {https://link.aps.org/doi/10.1103/PhysRevB.110.174515}
}

@Article{D4SC04465G,
author ="Novoa, Trinidad and di Mauro, Matias E. and Inostroza, Diego and El Haloui, Kaoutar and Sisourat, Nicolas and Maday, Yvon and Contreras-Garcia, Julia",
title  ="TcESTIME: Predicting High-Temperature Hydrogen-Based Superconductors",
journal  ="Chem. Sci.",
year  ="2025",
volume  ="16",
issue  ="1",
pages  ="57-68",
publisher  ="The Royal Society of Chemistry",
doi  ="10.1039/D4SC04465G",
url  ="http://dx.doi.org/10.1039/D4SC04465G"
}

@article{PhysRevB.101.144505,
  title = {Predicting Novel Superconducting Hydrides Using Machine Learning Approaches},
  author = {Hutcheon, Michael J. and Shipley, Alice M. and Needs, Richard J.},
  journal = {Phys. Rev. B},
  volume = {101},
  issue = {14},
  pages = {144505},
  numpages = {9},
  year = {2020},
  month = {Apr},
  publisher = {American Physical Society},
  doi = {10.1103/PhysRevB.101.144505},
  url = {https://link.aps.org/doi/10.1103/PhysRevB.101.144505}
}

@article{shipley2021,
  title = {High-throughput discovery of high-temperature conventional superconductors},
  author = {Shipley, Alice M. Hutcheon, M. J. and Needs, Richard J. and Pickard, C. J.},
  journal = {Phys. Rev. B},
  volume = {104},
  pages = {054501},
  year = {2021},
  publisher = {American Physical Society}
}

@article{belli2021strong,
  title={Strong Correlation Between Electronic Bonding Network and Critical Temperature in Hydrogen-Based Superconductors},
  author={Belli, Francesco and Novoa, Trinidad and Contreras-Garc{\'\i}a, Julia and Errea, Ion},
  journal={Nat. Commun.},
  volume={12},
  number={1},
  pages={5381},
  year={2021},
  publisher={Nature Publishing Group UK London}
}

@article{jiang2024data,
  title={Data-Driven Design of High-Temperature Superconductivity Among Ternary Hydrides Under Pressure},
  author={Jiang, Bowen and Luo, Xiaoshan and Iitaka, Toshiaki and Sun, Ying and Zhong, Xin and Lv, Jian and Xie, Yu and Ma, Yanming and Liu, Hanyu},
  journal={Phys. Rev. B}, 
  volume={111},
  pages={054505},
  year={2025},
}

@article{PhysRevLett.132.166001,
  title = {Feasible Route to High-Temperature Ambient-Pressure Hydride Superconductivity},
  author = {Dolui, Kapildeb and Conway, Lewis J. and Heil, Christoph and Strobel, Timothy A. and Prasankumar, Rohit P. and Pickard, Chris J.},
  journal = {Phys. Rev. Lett.},
  volume = {132},
  issue = {16},
  pages = {166001},
  numpages = {7},
  year = {2024},
  month = {Apr},
  publisher = {American Physical Society},
  doi = {10.1103/PhysRevLett.132.166001},
  url = {https://link.aps.org/doi/10.1103/PhysRevLett.132.166001}
}

@article{sanna2024prediction,
  title={Prediction of Ambient Pressure Conventional Superconductivity Above 80 K in Hydride Compounds},
  author={Sanna, Antonio and Cerqueira, Tiago FT and Fang, Yue-Wen and Errea, Ion and Ludwig, Alfred and Marques, Miguel AL},
  journal={npj Comput. Mater.},
  volume={10},
  number={1},
  pages={44},
  year={2024},
  publisher={Nature Publishing Group UK London}
}

@article{geng2022structural,
  title={Structural Diversity and Superconductivity in S-P-H Ternary Hydrides Under Pressure},
  author={Geng, Nisha and Bi, Tiange and Zurek, Eva},
  journal={J. Phy. Chem. C},
  volume={126},
  number={16},
  pages={7208-7220},
  year={2022},
  publisher={ACS Publications}
}

@article{yan2020metastable,
  title={A Metastable CaSH$_3$ Phase Composed of HS Honeycomb Sheets that is Superconducting Under Pressure},
  author={Yan, Yan and Bi, Tiange and Geng, Nisha and Wang, Xiaoyu and Zurek, Eva},
  journal={J. Phys. Chem. Lett.},
  volume={11},
  number={22},
  pages={9629-9636},
  year={2020},
  publisher={ACS Publications}
}

@article{PhysRevB.110.174508,
  title = {Stability-superconductivity Map for Compressed Na-intercalated Graphite},
  author = {Mishra, Shashi B. and Marcial, Edan T. and Debata, Suryakanti and Kolmogorov, Aleksey N. and Margine, Elena R.},
  journal = {Phys. Rev. B},
  volume = {110},
  issue = {17},
  pages = {174508},
  numpages = {13},
  year = {2024},
  month = {Nov},
  publisher = {American Physical Society},
  doi = {10.1103/PhysRevB.110.174508},
  url = {https://link.aps.org/doi/10.1103/PhysRevB.110.174508}
}

@article{geng2023conventional,
  title={Conventional High-Temperature Superconductivity in Metallic, Covalently Bonded, Binary-Guest C-B Clathrates},
  author={Geng, Nisha and Hilleke, Katerina P and Zhu, Li and Wang, Xiaoyu and Strobel, Timothy A and Zurek, Eva},
  journal={J. Am. Chem. Soc.},
  volume={145},
  number={3},
  pages={1696--1706},
  year={2023},
  publisher={ACS Publications}
}

@article{bassen2025real,
  title={Real-Space Orbital Tiling Approach for the Classification and Design of Superconductors},
  author={Bassen, Gregory and Bunstine, Wyatt and Han, Rebecca and Ebeid, Ragy and Zoghlin, Eli and McQueen, Tyrel M},
  journal={J. Phys. Condens. Matter},
  volume={37},
  number={45},
  pages={451501},
  year={2025},
  publisher={IOP Publishing}
}

@article{zhao2025prediction,
  title={Prediction of Pressure-Induced Superconductivity in the Ternary Systems CaYH 2n (n=3-6) at Moderate Pressures},
  author={Zhao, Jinghong and Chen, Bole and Li, Shichang and Chang, Ying and Yang, Xu and Chen, Mingyu and Li, Dengfeng},
  journal={J. Mater. Chem. C},
  volume={13},
  number={8},
  pages={4128--4136},
  year={2025},
  publisher={Royal Society of Chemistry}
}

@article{PhysRevB.73.024512,
  title = {Superconductivity Mediated by a Soft Phonon Mode: Specific Heat, Resistivity, Thermal Expansion, and Magnetization of $\mathrm{Y}{\mathrm{B}}_{6}$},
  author = {Lortz, R. and Wang, Y. and Tutsch, U. and Abe, S. and Meingast, C. and Popovich, P. and Knafo, W. and Shitsevalova, N. and Paderno, Yu. B. and Junod, A.},
  journal = {Phys. Rev. B},
  volume = {73},
  issue = {2},
  pages = {024512},
  numpages = {13},
  year = {2006},
  month = {Jan},
  publisher = {American Physical Society},
  doi = {10.1103/PhysRevB.73.024512},
  url = {https://link.aps.org/doi/10.1103/PhysRevB.73.024512}
}

@article{naik2023quantum,
  title={A Quantum-Chemical Bonding Database for Solid-State Materials},
  author={Naik, Aakash Ashok and Ertural, Christina and Dhamrait, Nidal and Benner, Philipp and George, Janine},
  journal={Sci. Data},
  volume={10},
  number={1},
  pages={610},
  year={2023},
  publisher={Nature Publishing Group UK London}
}

@article{hughbanks1983chains,
  title={Chains of Trans-Edge-Sharing Molybdenum Octahedra: Metal-Metal Bonding in Extended Systems},
  author={Hughbanks, Timothy and Hoffmann, Roald},
  journal={J. Am. Chem. Soc.},
  volume={105},
  number={11},
  pages={3528--3537},
  year={1983},
  publisher={ACS Publications}
}

@article{Liu:2017a,
  title = {Potential High-$T_{c}$ Superconducting Lanthanum and Yttrium Hydrides at High Pressure},
  author = {Liu, Hanyu and Naumov, Ivan I. and Hoffmann, Roald and Ashcroft, N. W. and Hemley, Russell J.},
  year = {2017},
  journal = {Proc. Natl. Acad. Sci. U. S. A.},
  volume = {114},
  number = {27},
  pages = {6990--6995},
  doi = {10.1073/pnas.1704505114},
  url = {https://doi.org/10.1073/pnas.1704505114}
}

@article{16r3-hjc7,
  title = {Wannier Functions Dually Localized in Space and Energy},
  author = {Mahler, Aaron and Williams, Jacob Z. and Su, Neil Qiang and Yang, Weitao},
  journal = {Phys. Rev. B},
  volume = {112},
  issue = {7},
  pages = {075137},
  numpages = {11},
  year = {2025},
  month = {Aug},
  publisher = {American Physical Society},
  doi = {10.1103/16r3-hjc7},
  url = {https://link.aps.org/doi/10.1103/16r3-hjc7}
}

@article{dronskowski1993crystal,
  title={Crystal Orbital Hamilton Populations (COHP): Energy-Resolved Visualization of Chemical Bonding in Solids Based on Density-Functional Calculations},
  author={Dronskowski, Richard and Bloechl, Peter E},
  journal={J. Phys. Chem.},
  volume={97},
  number={33},
  pages={8617--8624},
  year={1993},
  publisher={ACS Publications}
}

@article{RevModPhys.84.1419,
  title = {Maximally Localized Wannier Functions: Theory and Applications},
  author = {Marzari, Nicola and Mostofi, Arash A. and Yates, Jonathan R. and Souza, Ivo and Vanderbilt, David},
  journal = {Rev. Mod. Phys.},
  volume = {84},
  issue = {4},
  pages = {1419--1475},
  numpages = {0},
  year = {2012},
  month = {Oct},
  publisher = {American Physical Society},
  doi = {10.1103/RevModPhys.84.1419},
  url = {https://link.aps.org/doi/10.1103/RevModPhys.84.1419}
}

@article{muler2021crystal,
  title={Crystal Orbital Bond Index: Covalent Bond Orders in Solids},
  author={M\"uller, Peter C and Ertural, Christina and Hempelmann, Jan and Dronskowski, Richard},
  journal={J.Phys. Chem. C},
  volume={125},
  number={14},
  pages={7959--7970},
  year={2021},
  publisher={ACS Publications}
}

@article{ErreaSSCHA,
  title = { Anharmonic Free Energies and Phonon Dispersions From the Stochastic Self-Consisten Harmonic Approximation: Application to Platinum and Palladium Hydrides},
  author={Errea, Ion and Calandra, Matteo and Mauri, Francesco},
  journal={Phys. Rev. B},
  volume={89},
  number={6},
  pages={064302 (1-16)},
  year={2014},
  publisher={APS}
}

@incollection{cohen1994comments,
  title={Comments on the Maximum Superconducting Transition Temperature},
  author={Cohen, Marvin L and Anderson, PW},
  booktitle={A Career In Theoretical Physics},
  pages={287--298},
  year={1994},
  publisher={World Scientific}
}

@Inbook{OKeeffe1992,
author={O'Keeffe, Michael},
editor={Parth{\'e}, Erwin},
title={The Bond Valence Method in Crystal Chemistry},
bookTitle={Modern Perspectives in Inorganic Crystal Chemistry},
year={1992},
publisher={Springer Netherlands},
address={Dordrecht},
pages={163--175},
isbn={978-94-011-2726-4},
doi={10.1007/978-94-011-2726-4_9},
url={https://doi.org/10.1007/978-94-011-2726-4_9}
}

@article{savin1997elf,
  title={ELF: The Electron Localization Function},
  author={Savin, Andreas and Nesper, Reinhard and Wengert, Steffen and F{\"a}ssler, Thomas F},
  journal={Angew. Chem.},
  volume={36},
  number={17},
  pages={1808--1832},
  year={1997},
  publisher={Wiley Online Library}
}

@article{zurek2005muffin,
  title={Muffin-Tin Orbital Wannier-Like Functions for Insulators and Metals},
  author={Zurek, Eva and Jepsen, Ove and Andersen, Ole Krogh},
  journal={Chem. Phys. Chem.},
  volume={6},
  number={9},
  pages={1934--1942},
  year={2005},
  publisher={Wiley Online Library}
}

@article{PhysRev.52.191,
  title = {The Structure of Electronic Excitation Levels in Insulating Crystals},
  author = {Wannier, Gregory H.},
  journal = {Phys. Rev.},
  volume = {52},
  issue = {3},
  pages = {191--197},
  numpages = {0},
  year = {1937},
  month = {Aug},
  publisher = {American Physical Society},
  doi = {10.1103/PhysRev.52.191},
  url = {https://link.aps.org/doi/10.1103/PhysRev.52.191}
}

@article{radevych2025rigid,
  title={Rigid Muffin-tin Approximation in Plane-wave Codes for Fast Modeling of Phonon-mediated Superconductors},
  author={Radevych, Danylo and Shishidou, Tatsuya and Weinert, Michael and Margine, Elena R and Kolmogorov, Aleksey N and Mazin, Igor I},
  journal={arXiv preprint arXiv:2508.21281},
  year={2025}
}

@article{PhysRevLett.86.4366,
  title = {Superconductivity of ${\mathrm{MgB}}_{2}$: Covalent Bonds Driven Metallic},
  author = {An, J. M. and Pickett, W. E.},
  journal = {Phys. Rev. Lett.},
  volume = {86},
  issue = {19},
  pages = {4366--4369},
  numpages = {0},
  year = {2001},
  month = {May},
  publisher = {American Physical Society},
  doi = {10.1103/PhysRevLett.86.4366},
  url = {https://link.aps.org/doi/10.1103/PhysRevLett.86.4366}
}

@article{redington2024predicted,
  title={Predicted High-Pressure Hot Superconductivity in Li$_2$CaH$_{16}$ and Li2CaH17 Phases that Resemble the Type-II Clathrate Structure},
  author={Redington, Morgan and Zurek, Eva},
  journal={Chem. of Mater.},
  volume={36},
  number={17},
  pages={8412-8423},
  year={2024},
  publisher={ACS Publications}
}

@article{wang2024superconductivity,
  title={Superconductivity in Dilute Hydrides of Ammonia Under Pressure},
  author={Wang, Xiaoyu and Geng, Nisha and de Villa, Kyla and Militzer, Burkhard and Zurek, Eva},
  journal={J. Phys. Chem. Lett.},
  volume={15},
  number={22},
  pages={5947--5953},
  year={2024},
  publisher={ACS Publications}
}

@article{PhysRevB.12.905,
  title = {Transition Temperature of Strong-Coupled Superconductors Reanalyzed},
  author = {Allen, P. B. and Dynes, R. C.},
  journal = {Phys. Rev. B},
  volume = {12},
  issue = {3},
  pages = {905-922},
  numpages = {0},
  year = {1975},
  month = {Aug},
  publisher = {American Physical Society},
  doi = {10.1103/PhysRevB.12.905},
  url = {https://link.aps.org/doi/10.1103/PhysRevB.12.905}
}

@article{belli2025refining,
  title={Refining T$_\mathrm{c}$ Prediction in Hydrides via Symbolic-Regression-Enhanced Electron-Localization-Function-Based Descriptors},
  author={Belli, Francesco and Torres, Sean and Contreras-Garc{\'\i}a, Julia and Zurek, Eva},
  journal={Annalen der Physik},
  volume={537},
  number={11},
  pages={e00280},
  year={2025},
  publisher={Wiley Online Library}
}

@article{quinn2022identifying,
  title={Identifying New Classes of high Temperature Superconductors with Convolutional Neural Networks},
  author={Quinn, Margaret R and McQueen, Tyrel M},
  journal={Front. Electron. Mater.},
  volume={2},
  pages={893797},
  year={2022},
  publisher={Frontiers Media SA}
}

@article{Chu:1987a,
  author = {Wu, M. K. and Ashburn, J. R. and Torng, C. J. and Hor, P. H. and Meng, R. L. and Gao, L. and Huang, Z. J. and
            Wang, Y. Q. and Chu, C. W.},
  title = {Superconductivity at 93 K in a New Mixed-Phase Y-Ba-Cu-O Compound System at Ambient Pressure},
  journal = {Phys. Rev. Lett.},
  year = {1987}, 
  volume = {58},
  pages = {908-910}
}

@article{Gao:1994a,
  author = {Gao, L. and Xue, Y. Y. and Chen, F. and Xiong, Q. and Meng, R. L.
  and Ramirez, D. and Chu, C. W. and Eggert, J. H. and Mao, H. K.},
  title = {Superconductivity up to 164 K in HgBa$_2$Ca$_{m-1}$Cu$_m$O$_{2m+2+\delta}$ (m=1, 2, and 3) under quasihydrostatic pressures},
  journal = {Phys. Rev. B},
  year = {1994},
  volume = {50},
  pages = {4260 - 4263}
}

@article{bednorz1986possible,
  title={Possible High T$_c$ Superconductivity in the Ba-La Cu-O System},
  author={Bednorz, J George and M{\"u}ller, K Alex},
  journal={Z. Phys. B - Cond. Mat.},
  volume={64},
  number={2},
  pages={189--193},
  year={1986},
  publisher={Springer}
}

@ARTICLE{Laio:2002a,
  author = {Laio, A. and Parrinello, M.},
  title = {Escaping Free-energy Minima},
  journal = {Proc. Natl. Acad. Sci.},
  year = {2002},
  volume = {99},
  pages = {12562-12566}
}

@ARTICLE{highentropySC,
  author = {Zeng, L. and Li, L. adn Li, K. and Chen, R. and Luo, H.},
  title = {Recent Advances in High-Entropy Superconductors},
  journal = {npg Asia Materials},
  year = {2024}, 
  volume = {16},
  pages = {60}
}

@ARTICLE{nickelate,
  author = {Wang, Y. and Jiang, K. and Ying, J. and Wu, T. and Cheng, J. and Hu, J. and Chen, X.},
  title = {Recent Progress in Nickelate Superconductors},
  journal = {national Science Review},
  year = {2025}, 
  volume = {12},
  pages = {nwaf373}
}

@article{Godecker:2004,
        pages = {9911-9917 },
        year = {2004},
        author = {G\"odecker, S.},
        volume = {120},
        title = {Minima hopping: An efficient search method for the global minimum of the potential energy surface of complex molecular systems},
        journal =  {J. Chem. Phys.}
}

@article{jochen, 
        pages = {1032-1040},
        year = {2012},
        author = {Autschbach, J.},
        volume = {89},
        title = {Orbitals: Some Fiction and Some Facts},
        journal =  {J. Chem. Educ.}
}

@Article{Deng:pqp,
  author  = {Deng, L. and Habamahoro, T. and Safezoddeh, A. and Chu, C. W.},
  title   = {Ambient-pressure 151~K Superconductivity in {HgBa$_2$Ca$_2$Cu$_3$O$_{8+\delta}$} via Pressure Quench},
  journal = {Proc. Natl. Acad. Sci. U.S.A.}, 
  year    = {2026},
  volume  = {123},
  pages   = {e2536178123},
  url     = {https://doi.org/10.1073/pnas.2536178123},
}

@article{ponce,
  author =        {Zhang, Y. and Ponc\'{e}, S.},
  journal =       {arXiv preprint arXiv:608.04789},
  title =         {{In search of novel ductile superconductors}},
  year =          {2026},
  url =           {https://arxiv.org/abs/2608.04789},
}

@article{Wang:2012a,
  title = {Superconductive Sodalite-like Clathrate Calcium Hydride at High Pressures}, 
  author = {Wang, Hui and Tse, John S. and Tanaka, Kaori and Iitaka, Toshiaki and Ma, Yanming},
  year = {2012},
  journal = {Proc. Natl. Acad. Sci. U. S. A.},
  volume = {109},
  number = {17},
  pages = {6463--6466},
  doi = {10.1073/pnas.1118168109},
  url = {https://doi.org/10.1073/pnas.1118168109}
}

@article{Gao:2008a,
  author = {Gao, G. and Oganov, A. R. and Bergara, A. and Martinez-Canales, M. and Cui, T. and Iitaka, T. and Ma, Y. and Zou, G.},
  title = {Superconducting High Pressure Phase of Germane},
  journal = {Phys. Rev. Lett.},
  year = {2008},
  volume = {101},
  pages = {107002 (1-4)}
}

@InCollection{Zurek:2014d,
  author    = {Zurek, E.},
  title     = {Discovering New Materials via \emph{A Priori} Crystal Structure Prediction},
  booktitle = {Reviews in computational chemistry},
  publisher = {John Wiley \& Sons, Inc.},
  year      = {2016},
  editor    = {Parrill, A. L. and Lipkowitz, K. B.},
  volume    = {29},
  pages     = {274-326},
  address   = {Hoboken, New Jersey},
  doi       = {10.1002/9781119148739.ch5},
  url       = {https://doi.org/10.1002/9781119148739.ch5},
}

@Article{Zurek:2017c,
  author  = {Bi, T. and Miller, D. P. and Shamp, A. and Zurek, E.},
  title   = {Superconducting Phases of Phosphorus Hydride Under Pressure: Stabilization by Mobile Molecular Hydrogen},
  journal = {Angew. Chem. Int. Ed.},
  year    = {2017},
  volume  = {56},
  pages   = {10192-10195},
  doi     = {10.1002/ange.201701660},
  url     = {https://doi.org/10.1002/ange.201701660},
}

@Article{Zurek:2019b,
  author  = {Avery, P. and Wang, X. and Oses, C. and Gossett, E. and Proserpio, D. M. and Toher, C. and Curtarolo, S. and Zurek, E.},
  title   = {Predicting Superhard Materials via a Machine Learning Informed Evolutionary Structure Search},
  journal = {npj Comput. Mater.},
  year    = {2019},
  volume  = {5},
  pages   = {89 (1-11)},
  url     = {https://doi.org/10.1038/s41524-019-0226-8},
}

@Article{Zurek:2020i,
  author  = {Falls, Z. and Avery, P. and Wang, X. and Hilleke, K. P. and Zurek, E.},
  title   = {The XtalOpt Evolutionary Algorithm for Crystal Structure Prediction},
  journal = {J. Phys. Chem. C},
  year    = {2021},
  volume  = {125},
  pages   = {1601-1620},
  doi     = {10.1021/acs.jpcc.0c09531},
  url     = {https://doi.org/10.1021/acs.jpcc.0c09531},
}

@Article{Zurek:2021k,
  author = {Hilleke, K. P. and Zurek, E.},
  title  = {Tuning Chemical Precompression: Theoretical Design and Crystal Chemistry of Novel Hydrides in the Quest for Warm and Light Superconductivity at Ambient Pressures},
  year   = {2022},
  journal = {J. Appl. Phys.},
  volume = {131},
  pages = {070901 (1-19)},
  url    = {https://doi.org/10.1063/5.0077748},
}

@Article{Zurek:2023d,
  author = {Wang, B. and Hilleke, K. P. and Hajinazar, S. and Frapper, G. and Zurek, E.},
  title  = {Structurally Constrained Evolutionary Algorithm for the Discovery and Design of Metastable Phases},
  year   = {2023},
  journal   = {J. Chem. Theory Comput.},
  volume = {19},
  pages ={7960-7971},
  url    = {https://doi.org/10.1021/acs.jctc.3c00594}
}

@Article{Zurek:2023e,
  author = {Racioppi, S. and Storm, C. V. and McMahon, M. I. and Zurek, E.},
  title  = {On the Electride Nature of Na-hP4},
  year   = {2023},
  journal   = {Angew. Chem. Int. Ed.},
  volume = {135},
  pages ={e202310802 (1-5)},
  url    = {https://doi.org/10.1002/ange.202310802}
}

@article{Zurek:2024a,
  author = {Racioppi, S. and Saffarian-Deemyad, I. and Holle, W. and Belli, F. and Smith, J. S. and Kenney-Benson, C. and Ferry, R. and Zurek, E. and Deemyad, S.},
  title  = {Lithium's Low-Temperature Phase Transitions: Insights into Quantum Lattice Dynamics and Superconductivity},
  year   = {2025},
  journal   = {Phys. Rev. B.},
  volume = {111},
  pages ={054111},
  url    = {https://doi.org/10.1103/PhysRevB.111.054111}
}

@article{Zurek:2024e,
  author = {Hajinazar, S. and Zurek, E.},
  title  = {XtalOpt Version 13: Multi-Objective Evolutionary Search for Novel Functional Materials},
  year   = {2024},
  journal = {Comput. Phys. Commun.},
  volume  = {304},
  pages   = {109306},
  url     = {https://doi.org/10.1016/j.cpc.2024.109306}
}

@article{Zurek:2024j,
  author = {Racioppi, S. and De la Roza, A. O. and Hajinazar, S. and Zurek, E.},
  title  = {Powder X-Ray Diffraction Assisted Evolutionary Algorithm for Crystal Structure Prediction},
  year   = {2025},
  volume = {4},
  pages = {73-83},
  journal = {Digital Discovery},
  url = {https://doi.org/10.1039/D4DD00269E},
}

@article{Zurek:2024p,
  title = {High-Pressure Electrides: A Quantum Chemical Perspective},
  author  = {Racioppi, S. and Zurek, E.},
  year   = {2025},
  journal = {Annu. Rev. Condens. Matter Phys.},
  volume = {55},
  pages = {421-442},
  url =  {https://doi.org/10.1146/annurev-matsci-080423-124050}
}

@article{Zurek:2025n,
  author = {Hajinazar, S. and Zurek, E.},
  title  = {XtalOpt Version 14: Variable-Composition Crystal Structure Search for Functional Materials Through Pareto Optimization},
  year   = {2026},
  journal = {Comput. Phys. Commun.},
  volume = {320},
  pages = {109910},
  url   = {https://doi.org/10.1016/j.cpc.2025.109910}
}

@article{Zurek:2025p,
  author = {Racioppi, S. and Zurek, E.},
  title  = {Using Topology to Predict Electrides in the Solid State},
  year   = {2025},
  journal = {J. Phys. Chem. A},
  volume = {129},
  pages = {10031-10038},
  url   = {https://doi.org/10.1021/acs.jpca.5c05317}
}

@article{Zurek:2025q,
  author = {Prasankumar, R. P. and Julian, M. and Hutcheon, M. and Heil, C. and Deng, L. and Basov, D. and Chu, C.-W. and Comin, R. and Kim, P. and Meredig, B. and Pickard, C. and Pickett, W. and Strobel, T. and Wolf, S. and Zurek, E. and Myhrvold, N.},
  title  = {The Path to Room-Temperature Superconductivity: A Programmatic Approach},
  year   = {2026},
  journal = {Proc. Natl. Acad. Sci. U.S.A.},
  volume = {123},
  pages = {e2520324123},
  url   = {https://doi.org/10.1073/pnas.2520324123}
}

@article{Zurek:2026e,
  title = {Metal Atom (Dis)Order and Superconductivity in YCaH$_n$ ($n=8-20)$ High-Pressure Superhydrides},
  author  = {Kimura, M. W. and Jang, S. W. and Geng, N. and Zurek, E.},
  year   = {2026},
  journal = {Inorg. Chem.},
  volume = {65},
  pages = {17538-17551},
  url   = {https://doi.org/10.1021/acs.inorgchem.6c02106}
}

@article{Zurek:2026o,
  title = {Anharmonicity and nonadiabaticity in hydride superconductors},
  author  = {Mishra, S. B. and Belli, F. and Zurek, E. and Margine, E. R.},
  year   = {2026},
  note = {https://doi.org/10.48550/arXiv.2608.14949}
}

\end{document}